\documentclass[journal, onecolumn,12pt, draftclsnofoot]{IEEEtran}

\usepackage{cleveref}  % used to ref propositions

\usepackage{mathrsfs}

\usepackage[noadjust]{cite}
\usepackage{graphicx,color,overpic,psfrag}
\usepackage{amsmath, amssymb}
\usepackage{latexsym}
\usepackage{bm}
\usepackage{amssymb}
\usepackage{cases}
\usepackage{array}
\usepackage{fancyhdr}
\usepackage{setspace}

\ifCLASSOPTIONcompsoc
\usepackage[caption=false,font=normalsize,labelfont=sf,textfont=sf]{subfig}
\else
\usepackage[caption=false,font=footnotesize]{subfig}
\fi

\usepackage{url}
\usepackage{algpseudocode}
\usepackage{algorithm}
\usepackage{blkarray}
\usepackage{booktabs}

\usepackage{multirow}
\usepackage{dsfont}
\usepackage{tabularx}
\usepackage[table]{xcolor}

\usepackage{amsfonts}

\usepackage{letltxmacro}
\graphicspath{{figure/}}%path of the figures

\newtheorem{remark}{Remark}

\newtheorem{prop}{Proposition}

\newcommand{\figref}[1]{Fig. \ref{#1}}
\newcommand{\tabref}[1]{Table \ref{#1}}

\newcommand{\appref}[1]{Appendix \ref{#1}}
\newcommand{\secref}[1]{Section \ref{#1}}

\newcommand{\propref}[1]{Proposition \ref{#1}}

\newcommand{\Exp}{{\mathsf{E}}}
\newcommand{\expect}[1]{\Exp\left\{#1\right\}}

\newcommand{\diag}[1]{\mathsf{diag}\left\{#1\right\}}

\newcommand{\abs}[1]{\left|#1\right|}
\newcommand{\sqrabs}[1]{\left|#1\right|^{2}}
\newcommand{\normt}[1]{\left\|#1\right\|}
\newcommand{\sqrnormt}[1]{\left\|#1\right\|^{2}}
\newcommand{\oneon}[1]{\frac{1}{#1}}
\newcommand{\intd}{\mathrm{d}}
\newcommand{\intdx}[1]{\intd #1}

\newcommand{\GCN}[2]{\mathcal{CN}\left( #1 , #2\right) }
\newcommand{\ith}[1]{$#1$th}
\newcommand{\delfunc}[1]{\delta\left(#1\right)}

\newcommand{\vecele}[2]{\left[#1\right]_{#2}}
\newcommand{\thetabs}[2]{{\dnnot{\theta}{bs}}}

\newcommand{\parenth}[1]{\left(#1\right)}

\newcommand{\equaa}{\mathop{=}^{(\mathrm{a})}}
\newcommand{\equab}{\mathop{=}^{(\mathrm{b})}}
\newcommand{\equac}{\mathop{=}^{(\mathrm{c})}}

\newcommand{\cK}{\mathcal{K}}

\newcommand{\bb}{\mathbf{b}}

\newcommand{\be}{\mathbf{e}}

\newcommand{\bg}{\mathbf{g}}

\newcommand{\bu}{\mathbf{u}}
\newcommand{\bv}{\mathbf{v}}
\newcommand{\bw}{\mathbf{w}}
\newcommand{\bx}{\mathbf{x}}
\newcommand{\by}{\mathbf{y}}
\newcommand{\bz}{\mathbf{z}}

\newcommand{\bA}{\mathbf{A}}
\newcommand{\bB}{\mathbf{B}}

\newcommand{\bI}{\mathbf{I}}

\newcommand{\bV}{\mathbf{V}}

\newcommand{\bX}{\mathbf{X}}

\newcommand{\bbC}{\mathbb{C}}

\newcommand{\bzero}{\mathbf{0}}

\newcommand{\expx}[1]{\exp\left\{#1\right\}}

\newcommand{\upnot}[2]{#1^{\mathrm{#2}}}
\newcommand{\dnnot}[2]{#1_{\mathrm{#2}}}

\newcommand{\barjmath}{\bar{\jmath}}

\newcommand{\ntb}{\notag\\}

\newcommand{\DL}{\mathrm{dl}}
\newcommand{\MAX}{\mathrm{max}}
\newcommand{\MIN}{\mathrm{min}}
\newcommand{\ASLNR}{\mathrm{aslnr}}

\newcommand{\udbdb}{\underline{\mathbf{b}}}
\newcommand{\bdGamma}{\boldsymbol{\Gamma}}

\allowdisplaybreaks

\begin{document}

\title{Massive MIMO Transmission for\\ LEO Satellite Communications}

\author{

Li~You, Ke-Xin~Li,
Jiaheng~Wang,
Xiqi~Gao, Xiang-Gen~Xia,
and~Bj\"{o}rn~Ottersten%

\thanks{This work will be presented in part at the IEEE International Conference on Communications, Dublin, Ireland, Jun. 2020 \cite{You20Massive}.
}% <-this % stops a space
\thanks{
L. You, K.-X. Li, J. Wang, and X. Q. Gao are with the National Mobile Communications Research Laboratory, Southeast University, Nanjing 210096, China, and also with the Purple Mountain Laboratories, Nanjing 211100, China (e-mail: liyou@seu.edu.cn; likexin3488@seu.edu.cn; jhwang@seu.edu.cn; xqgao@seu.edu.cn).
X.-G. Xia is with the Department of Electrical and Computer Engineering, University of Delaware, Newark, DE 19716 USA (e-mail: xxia@ee.udel.edu).
B. Ottersten is with the Interdisciplinary Centre for Security, Reliability and Trust (SnT), University of Luxembourg, L-2721 Luxembourg City, Luxembourg (e-mail: bjorn.ottersten@uni.lu). \emph{(Corresponding author: Xiqi Gao.)}
}
}

\maketitle

\begin{abstract}
\begin{spacing}{1.0}%%行间距
Low earth orbit (LEO) satellite communications are expected to be incorporated in future wireless networks, in particular 5G and beyond networks, to provide global wireless access with enhanced data rates.
Massive multiple-input multiple-output (MIMO) techniques, though widely used in terrestrial communication systems, have not been applied to LEO satellite communication systems.
In this paper, we propose a massive MIMO transmission scheme with full frequency
reuse (FFR) for LEO satellite communication systems and exploit statistical channel state information (sCSI) to address the difficulty of obtaining instantaneous CSI (iCSI) at the transmitter.
We first establish the massive MIMO channel model for LEO satellite communications and simplify the transmission designs via performing Doppler and delay compensations at user terminals (UTs).
Then, we develop the low-complexity sCSI based downlink (DL) precoder and uplink (UL) receiver in closed-form, aiming to maximize the average signal-to-leakage-plus-noise ratio (ASLNR) and the average signal-to-interference-plus-noise ratio (ASINR), respectively.
It is shown that the DL ASLNRs and UL ASINRs of all UTs reach their upper bounds under some channel condition. Motivated by this, we propose a space angle based user grouping (SAUG) algorithm to schedule the served UTs into different groups, where each group of UTs use the same time and frequency resource. The proposed algorithm is asymptotically optimal in the sense that the lower and upper bounds of the achievable rate coincide when the number of satellite antennas or UT groups is sufficiently large.
Numerical results demonstrate that the proposed massive MIMO transmission scheme with FFR significantly enhances the data rate of LEO satellite communication systems. Notably, the proposed sCSI based precoder and receiver achieve the similar performance with the iCSI based ones that are often infeasible in practice.
\end{spacing}
\end{abstract}
\begin{IEEEkeywords}
LEO satellite, massive MIMO, multibeam satellite, full frequency reuse, statistical CSI, user grouping.
\end{IEEEkeywords}

\newpage

\section{Introduction}\label{sec:net_intro}

Satellite communication systems can provide seamless wireless coverage so as to complement and extend terrestrial communication networks and, as in recent standardization endeavors \cite{Guidotti19Architectures}, are expected to be incorporated in future wireless networks, in particular 5G and beyond networks. Low earth orbit (LEO) satellite communications, with orbits at altitudes of less than 2000 km, have recently gained broad research interests due to the potential in providing global wireless access with enhanced data rates. Compared with the geostationary earth orbit (GEO) counterpart, LEO satellite communication systems impose much less stringent requirements on, e.g., power consumption and transmission signal delays. Recently, several projects, e.g., OneWeb and SpaceX, on LEO satellite communication systems have been launched \cite{Di19Ultra}.

In satellite communication systems, multibeam transmission techniques have been widely adopted to increase transmission data rates. As a well-know multibeam solution, a four-color frequency reuse (FR4) scheme where adjacent beams are allocated with non-overlapping frequency spectrum (or different polarizations) is adopted to mitigate the co-channel inter-beam interference \cite{Schwarz19MIMO,Vazquez16Precoding}. To further enhance the spectral efficiency of satellite communications, the more aggressive full frequency reuse (FFR) schemes \cite{Schwarz19MIMO,Vazquez16Precoding,Letzepis08Capacity,Wang18Robust,You19Outage}, where frequency resources are reused across neighboring beams, have been considered to increase the total available bandwidth in each beam as that has been done in terrestrial cellular systems. Yet, in FFR the inter-beam interference becomes a critical issue, which has to be properly handled. In general, inter-beam interference management can be performed at either the transmitter via precoding or at the receiver via multi-user detection, similar as in terrestrial cellular communication systems \cite{Yoo11Common}. Compared with non-linear dirty paper coding (DPC) precoding and multi-user detection, in practice linear precoding and detection are more preferred in multibeam satellite communication systems due to their low computational complexity and near-optimal performance \cite{Lee07High}.

It is worth noting that most of the existing works on downlink (DL) precoding in multibeam satellite communications, e.g., \cite{Schwarz19MIMO,Vazquez16Precoding}, rely on precise instantaneous channel state information (iCSI).
However, obtaining iCSI at the transmitter sides of satellite communication systems is usually difficult and even infeasible due to a number of practical factors, especially the long propagation delay between a satellite and user terminals (UTs) as well as the mobility of UTs and satellites. In particular, for time-division duplex (TDD) systems, the coherence time of the channel is shorter than the transmission delay, which makes obtaining accurate iCSI via the UL-DL reciprocity a mission impossible. On the other hand, in more common frequency-division duplex (FDD) systems, obtaining iCSI at the satellite side requires UL feedback from UTs, which inevitably introduces a great among of training and feedback overhead due to mobility of UTs and more importantly could become outdated as a result of the long propagation delay.

In recent years, massive multiple-input multiple-output (MIMO) transmission, where a large number of antennas are equipped at a base station to serve many UTs, has been applied in terrestrial cellular wireless networks, e.g., 5G \cite{Marzetta10Noncooperative,Lu14overview}, as an enabling technology. Massive MIMO can substantially increase available degrees of freedom, enhance spectral efficiency, and achieve high data rates. Motivated by this, we propose to exploit massive MIMO along with FFR for LEO satellite communication systems, where a large number of antennas are equipped at the LEO satellite side. Our focus is particularized on the physical layer transmission design for massive MIMO LEO satellite communication systems.
We note that it is not necessary to perform predefined multiple beamforming in fully digital-implemented FFR satellite communication systems. Exploiting massive MIMO for satellite communications with FFR can be seen as a technique without predefined beamforming.

Albeit the existence of a large body of literature on massive MIMO in terrestrial cellular communication systems \cite{Lu14overview}, so far massive MIMO has not been applied to satellite communication systems.
The performance of massive MIMO systems relies substantially on the available CSI \cite{You15Pilot,Sun15Beam,Lu2019Robust}.
As mentioned above, obtaining accurate DL iCSI at the LEO satellite side is generally difficult and even infeasible due to the long propagation delay and the mobility of satellites and UTs,
which makes inapplicable of the existing terrestrial massive MIMO transmission approaches relying on iCSI.
Meanwhile, the implementation complexity accompanied with massive MIMO becomes a critical concern in satellite communication systems considering the payload limitation on satellites.
Consequently, incorporating massive MIMO into LEO satellite communication systems is still an open and challenging task.

For massive MIMO, obtaining iCSI at the transmitter has been a difficult problem even in terrestrial communication systems, especially in high-mobility scenarios.
Compared with iCSI, statistical CSI (sCSI) varies much slower and thus can be relatively easily obtained at both the satellite and the UTs with sufficiently high accuracy.
Hence, sCSI based DL precoding has been proposed in terrestrial massive MIMO systems \cite{Sun15Beam,You18Non,Lu2019Robust}. For massive MIMO satellite communication systems, it is more practical to use sCSI, which can overcome the difficulty of acquiring iCSI and significantly reduce the computational overhead of satellite payloads via the much less frequent update of transmission strategies including, e.g., DL precoding, UL receiving, and user grouping.

In this paper, we investigate massive MIMO transmission for LEO satellite communication systems using FFR based on sCSI. In particular, we focus on devising DL precoding, UL receiving, and user grouping utilizing sCSI. While this paper focuses on the LEO satellite communications, the proposed massive MIMO transmission schemes can also be extended to other non-terrestrial communication systems, e.g., GEO satellite communication systems, and high-altitude platform (HAP) communication systems. The major contributions of the current work are summarized as follows:
\begin{itemize}
\item We introduce massive MIMO into LEO satellite communication systems using FFR and investigate low-complexity and low-overhead transmission strategies based on sCSI.
\item We establish the massive MIMO channel model for LEO satellite communications by incorporating the LEO satellite signal propagation properties, and simplify the UL/DL transmission designs via performing Doppler and delay compensations at UTs.
\item We develop the sCSI based DL precoder and UL receiver in closed-form, aiming to maximize the average signal-to-leakage-plus-noise ratio (ASLNR) and the average signal-to-interference-plus-noise ratio (ASINR), respectively, and theoretically prove that the proposed sCSI based scheme asymptotically approaches the iCSI based one.
\item We propose a space angle based user grouping (SAUG) algorithm using only the channel space angle information, and show that the proposed algorithm is asymptotically optimal in the sense that the lower and upper bounds of the achievable rate coincide when the number of satellite antennas or UT groups is sufficiently large.
\item Simulation results demonstrate that the proposed massive MIMO transmission scheme with FFR significantly enhances the data rate of LEO satellite communication systems. Notably, the proposed sCSI based precoder and receiver achieve the similar performance with the iCSI based ones that are often infeasible in practice.
\end{itemize}

The rest of the paper is organized as follows. In \secref{sec:sys_mod}, we investigate the channel model and the corresponding transmission signal model for LEO satellite communication systems. Based on the system model, we then investigate the optimal DL and UL transmission strategies for LEO satellite communications in \secref{sec:ud_trans}.
In \secref{sec:ut_sc}, we further investigate user grouping. We present the numerical results in \secref{sec:sim_n} and conclude the paper in \secref{sec:conc}.
The major variables adopted in the paper is listed in \tabref{notation_list} for ease of reference.

\newcolumntype{L}{>{\hspace*{-\tabcolsep}}l}
\newcolumntype{R}{l<{\hspace*{-\tabcolsep}}}
\definecolor{lightblue}{rgb}{0.93,0.95,1.0}
\begin{table}[!t]
	\centering
	\renewcommand\arraystretch{1.2}
	\footnotesize
	\caption{Variable List}
	\label{notation_list}
	\begin{tabular}{LcR}
		\toprule
		Notation && Definition 				\\
		\midrule
		\rowcolor{lightblue}
		$\dnnot{M}{x}$, $\dnnot{M}{y}$	&& Numbers of antennas of UPA at x- and y- axes 		\\
		$\upnot{g}{dl}_{k,p}$, $\upnot{g}{ul}_{k,p}$  && Complex channel gains for DL and UL		\\
		\rowcolor{lightblue}
		$P_k$  && Number of multipaths \\
		$\nu_{k,p}$, $\upnot{\nu}{sat}_{k,p}$, $\upnot{\nu}{sat}_{k}$, $\upnot{\nu}{ut}_{k,p}$ && Doppler frequencies \\
		\rowcolor{lightblue}
		$\tau_{k,p}$, $\upnot{\tau}{min}_{k}$, $\upnot{\tau}{max}_{k}$, $\upnot{\tau}{ut}_{k,p}$  &&  Propagation delays \\
		$\bv_{k,p}$, $\bv_{k,p}^\mathrm{x}$, $\bv_{k,p}^\mathrm{y}$, $\bv_{k}$ && DL array response vectors \\
		\rowcolor{lightblue}
		$\bu_{k}$, $\bu_{k}^\mathrm{x}$, $\bu_{k}^\mathrm{y}$ && UL array response vectors \\
		$\upnot{\bg}{dl}_{k}\left(t,f\right)$, $\upnot{\bg}{ul}_{k}\left(t,f\right)$ && DL and UL channel vectors \\
		\rowcolor{lightblue}
		$\upnot{g}{dl}_{k}\left(t,f\right)$, $\upnot{g}{ul}_{k}\left(t,f\right)$ && Complex channel gains after compensation \\
		$\theta_{k,p}^{\mathrm{x}}$, $\theta_{k,p}^{\mathrm{y}}$, $\theta_{k}^{\mathrm{x}}$, $\theta_{k}^{\mathrm{y}}$ && Angles \\
		\rowcolor{lightblue}
		 $\vartheta_{k,p}^{\mathrm{x}}$, $\vartheta_{k,p}^{\mathrm{y}}$, $\vartheta_{k}^{\mathrm{x}}$, $\vartheta_{k}^{\mathrm{y}}$ && Space angles \\
		$\gamma_{k}$ && Channel power \\
		\rowcolor{lightblue}
		$\kappa_{k}$ && Rician factor \\
		$\dnnot{N}{us}$, $\dnnot{N}{cp}$ && Numbers of subcarriers and CP \\
		\rowcolor{lightblue}
		$\dnnot{T}{s}$ && System sampling interval \\
		$\dnnot{T}{us}$, $\dnnot{T}{cp}$ && Lengths of OFDM symbol and CP \\
		\rowcolor{lightblue}
		$\upnot{\bg}{dl}_{k,\ell,n}$, $\upnot{\bg}{ul}_{k,\ell,n}$ && Effective DL and UL frequency domain channel vectors after compensation \\
		$\upnot{g}{dl}_{k,\ell,n}$, $\upnot{g}{ul}_{k,\ell,n}$ && Effective DL and UL frequency domain channel gains after compensation \\
		\rowcolor{lightblue}
		$\upnot{\bg}{dl}_{k}$, $\upnot{\bg}{ul}_{k}$ && DL and UL channel vectors \\
		$\upnot{q}{dl}_{k}$, $\upnot{q}{ul}$ && DL and UL transmit power \\
		\rowcolor{lightblue}
		$\bb_{k}$, $\bw_{k}$ && Normalized DL precoder and UL receiver \\
		$\mathsf{ASLNR}_{k}$, $\mathsf{ASINR}_{k}$ && DL ASLNR and UL ASINR \\
		\rowcolor{lightblue}
		$G_{\mathrm{x}}$, $G_{\mathrm{y}}$ && Number of groups at x- and y- axes \\
		$\Delta_{\mathrm{x}}$, $\Delta_{\mathrm{y}}$ && Lengths of space angle interval at x- and y- axes \\
		\rowcolor{lightblue}
		$\mathcal{A}_{\left( g,r \right)}^{\left(m,n\right)}$ && Space angle interval \\
		$\mathcal{K}_{\left(g,r\right)}$ && Set of UTs in group $(g,r)$ \\
		\rowcolor{lightblue}
		$\dnnot{R}{dl}$, $\dnnot{R}{dl}^{\mathrm{ub}}$, $\dnnot{R}{dl}^{\mathrm{lb}}$ && DL ergodic rate and its upper bound, lower bound \\
		\bottomrule
	\end{tabular}	
\end{table}

\section{System Model}\label{sec:sys_mod}

\subsection{System Setup}

\begin{figure}[!t]
 \centering
 \includegraphics[width=4.1in]{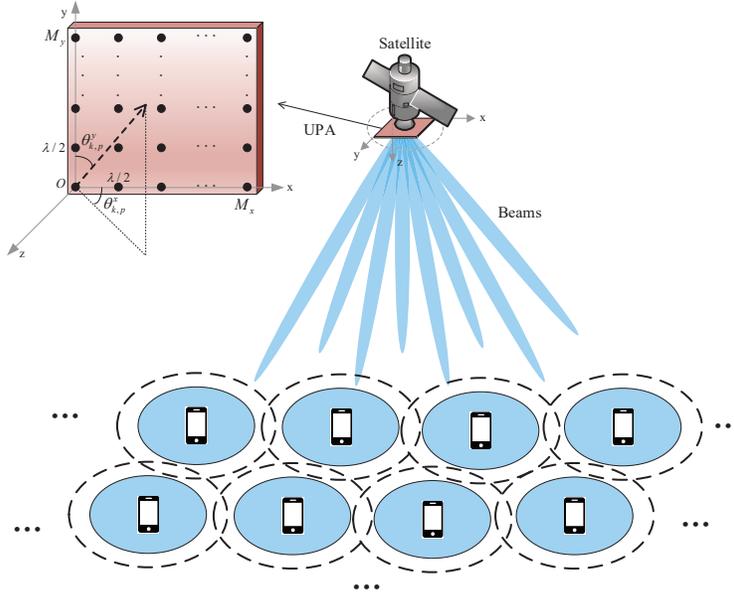}
 \caption{Illustration of the LEO satellite communication system setup.}
 \label{fig_LEO_satellite_system}
\end{figure}

Consider a LEO satellite communication system where a satellite provides services to a number of single-antenna UTs simultaneously. The satellite is equipped with a uniform planar array (UPA) composed of $M=\dnnot{M}{x}\dnnot{M}{y}$ antennas where $\dnnot{M}{x}$ and $\dnnot{M}{y}$ are the numbers of antennas on the x- and y-axes, respectively. Assume without loss of generality that the antennas are separated by one-half wavelength in both the x- and y-axes, and both $\dnnot{M}{x}$ and $\dnnot{M}{y}$ are even. The system setup is illustrated in \figref{fig_LEO_satellite_system}.

\subsection{DL Channel Model}

As different UTs are usually spatially separated by a few wavelengths, it is reasonable to assume that the channel realizations between the satellite and different UTs are uncorrelated \cite{Adhikary13Joint}. We focus on investigating the DL channel between the satellite and UT $k$.
Using a ray-tracing based channel modeling approach, the complex baseband DL space domain channel response between the LEO satellite and UT $k$ at instant $t$ and frequency $f$ can be represented by \cite{Kanatas16Radio,Clerckx13MIMO,You16Channel}
\begin{align}\label{eq:sat_cha_mod_ray}
  \upnot{\bg}{dl}_{k}\left(t,f\right)=\sum_{p=0}^{P_{k}-1}\upnot{g}{dl}_{k,p}
  \cdot\expx{\barjmath2\pi\left[t\nu_{k,p}-f\tau_{k,p}\right]}
  \cdot\bv_{k,p}
  \in\bbC^{M\times 1},
\end{align}
where $\bbC^{M\times N}$ denotes the $M\times N$ dimensional complex-valued vector space, $\barjmath=\sqrt{-1}$, $P_{k}$ denotes the number of channel propagation paths of UT $k$, and $\upnot{g}{dl}_{k,p}$, $\nu_{k,p}$, $\tau_{k,p}$, and $\bv_{k,p}\in\bbC^{M\times 1}$ are the complex-valued gain, the Doppler shift, the propagation delay, and the DL array response vector associated with path $p$ of UT $k$, respectively.
Note that the channel model adopted in \eqref{eq:sat_cha_mod_ray} is applicable over the time intervals of interest where the relative positions of the LEO satellite and UT $k$ do not change significantly, and thus the physical channel parameters, $P_{k}$, $\upnot{g}{dl}_{k,p}$, $\nu_{k,p}$, $\tau_{k,p}$, and $\bv_{k,p}$, are assumed to be invariant. When the LEO satellite and/or the UT move over large distances, the above channel parameters will vary and should be updated accordingly \cite{Clerckx13MIMO}.
It is worth mentioning that the ray-tracing based channel model in \eqref{eq:sat_cha_mod_ray} can be applied to different propagation scenarios, and further analysis of the channel model will depend on the parameter properties in the specific scenario.
Hereafter, we detail some propagation characteristics of the LEO satellite channels and their impact on the modeling of the channel parameters in \eqref{eq:sat_cha_mod_ray}.

\subsubsection{Doppler}

For LEO satellite communications, assuming that the scatterers are stationary in the considered interval of interest, then the Doppler shift $\nu_{k,p}$ associated with propagation path $p$ of UT $k$ is mainly composed of two independent Doppler shifts, $\upnot{\nu}{sat}_{k,p}$ and $\upnot{\nu}{ut}_{k,p}$, that are caused by the motions of the LEO satellite and the UT, respectively \cite{Papathanassiou01comparison,Perez01Statistical}.

It is worth noting that due to the relatively high altitude of the LEO satellite, the Doppler shifts $\upnot{\nu}{sat}_{k,p}$ caused by the motion of the LEO satellite can be assumed to be identical for different propagation paths $p$ of the same UT $k$ \cite{Papathanassiou01comparison,Perez01Statistical}, and different for different UTs. Thus, for notation simplicity, we omit the path index of the Doppler shift $\upnot{\nu}{sat}_{k,p}$ due to the motion of the LEO satellite and rewrite the Doppler shifts as $\upnot{\nu}{sat}_{k,p}=\upnot{\nu}{sat}_{k}$.
On the other hand, the Doppler shifts $\upnot{\nu}{ut}_{k,p}$ due to the motion of the UT are typically different for different propagation paths, which contribute the Doppler spread of the LEO satellite channels \cite{Papathanassiou01comparison,Perez01Statistical}. As the scattering characteristics around the UTs mainly determine the Doppler shifts caused by the movement of the UTs, the modeling of the Doppler spread in LEO satellite communications can be similar to that in the traditional terrestrial cellular communications \cite{Perez01Statistical}.

\subsubsection{Delay}

Due to the relatively large distance between the LEO satellite and the UTs, the propagation delay $\tau_{k,p}$ associated with path $p$ of UT $k$ exhibits a much larger value than that in terrestrial wireless channels. Denote by $\upnot{\tau}{min}_{k}=\min_{p}\left\{\tau_{k,p}\right\}$ and $\upnot{\tau}{max}_{k}=\max_{p}\left\{\tau_{k,p}\right\}$ the minimum and maximum values of the propagation delays of UT $k$, respectively.
The delay spread of the LEO satellite channels $\upnot{\tau}{max}_{k}-\upnot{\tau}{min}_{k}$ might be much smaller than that of the terrestrial wireless channels as observed in measurement results \cite{Vojcic94Performance,3gpp.38.811,Perez01Statistical}.
For notational brevity, we define $\upnot{\tau}{ut}_{k,p}\triangleq\tau_{k,p}-\upnot{\tau}{min}_{k}$.
Note that due to, e.g., the long propagation delays in LEO satellite communications, acquiring reliable iCSI at the transmitter sides is usually infeasible, especially when the UTs are in high mobility. Thus, it is more practical to investigate transmission design with, e.g., sCSI, in LEO satellite communications.

\subsubsection{Angle}

The UPA response vector $\bv_{k,p}$ in \eqref{eq:sat_cha_mod_ray} can be represented by \cite{Balanis16Antenna,Shafin17Angle}
\begin{align}\label{eq:arr_resp_sat_mat}
\bv_{k,p}&\triangleq\bv_{k,p}^\mathrm{x}\otimes\bv_{k,p}^\mathrm{y}\ntb
&=\bv_{\mathrm{x}}\left(\vartheta_{k,p}^\mathrm{x}\right)\otimes
\bv_{\mathrm{y}}\left(\vartheta_{k,p}^\mathrm{y}\right)
\in\bbC^{M\times 1},
\end{align}
where $\otimes$ denotes the Kronecker product, and $\bv_{k,p}^{d}$ for $d\in\mathcal{D}\triangleq\left\{\mathrm{x},\mathrm{y}\right\}$ is the array response vector of the angle with respect to the x- or y-axis given by
\begin{align}\label{eq:ula steer dm}
\bv_{k,p}^{d}
&\triangleq
\bv_{d}\left(\vartheta_{k,p}^{d}\right) \ntb
&=\oneon{\sqrt{M_{d}}}\left[1\ \expx{-\barjmath\pi\vartheta_{k,p}^{d}}
\ \ldots\ \expx{-\barjmath\pi(M_{d}-1)\vartheta_{k,p}^{d}}\right]^{T}\in\bbC^{M_{d}\times 1},
\end{align}
with the superscript $(\cdot)^{T}$ denoting the transpose operation.
In \eqref{eq:ula steer dm}, the parameters $\vartheta_{k,p}^{\mathrm{x}}$ and $\vartheta_{k,p}^{\mathrm{y}}$ are related to the physical angles as $\vartheta_{k,p}^{\mathrm{x}}=\sin\left(\theta_{k,p}^{\mathrm{y}}\right)\cos\left(\theta_{k,p}^{\mathrm{x}}\right)$ and $\vartheta_{k,p}^{\mathrm{y}}=\cos\left(\theta_{k,p}^{\mathrm{y}}\right)$ where
$\theta_{k,p}^{\mathrm{x}}$ and $\theta_{k,p}^{\mathrm{y}}$ are the angles with respect to the x- and y-axes associated with the \ith{p} propagation path of UT $k$, respectively.
For satellite communication channels, the angles of all propagation paths associated with the same UT, can be assumed to be identical due to the relatively high altitude of the satellite compared with that of the scatterers located in the vicinity of the UTs \cite{Jaeckel17QuaDRiGa}, i.e., $\vartheta_{k,p}^{d}=\vartheta_{k}^{d}$. Note that the parameters $\vartheta_{k}^{d}$ can reflect the propagation properties of the LEO satellite channels in the space domain, and we refer to $\vartheta_{k}^{d}$ as the space angle parameters. Then, the array response vector can be rewritten as
\begin{align}\label{eq:arr_resp_sing}
\bv_{k,p}=\bv_{k}&=\bv_{k}^\mathrm{x}\otimes\bv_{k}^\mathrm{y} \ntb
&=\bv_{\mathrm{x}}\left(\vartheta_{k}^\mathrm{x}\right)\otimes
\bv_{\mathrm{y}}\left(\vartheta_{k}^\mathrm{y}\right)
\in\bbC^{M\times 1},
\end{align}
which will be referred to as the DL channel direction vector of UT $k$ that is associated with the space angles $\vartheta_{k}^\mathrm{x}$ and $\vartheta_{k}^\mathrm{y}$.
Note that when the number of antennas $M_{d}$ for $d\in\mathcal{D}$ tends to infinity, we can know from \eqref{eq:ula steer dm} and \eqref{eq:arr_resp_sing} that the channel direction vectors of different UTs are asymptotically orthogonal, i.e.,
\begin{align}\label{eq:asy_ort_dl}
  \lim_{M_{d}\to\infty}\left(\bv_{k}^{d}\right)^{H}\bv_{k'}^{d}=\delfunc{k-k'},
\end{align}
where $(\cdot)^{H}$ denotes the conjugate-transpose operation.

Based on the above modeling of the propagation properties of LEO satellite communications, we can rewrite the channel response in \eqref{eq:sat_cha_mod_ray} as follows
\begin{align}\label{eq:sat_cha_mod_ray_re}
  \upnot{\bg}{dl}_{k}\left(t,f\right)
  =\expx{\barjmath2\pi\left[t\upnot{\nu}{sat}_{k}-f\upnot{\tau}{min}_{k}\right]}
  \cdot\upnot{g}{dl}_{k}\left(t,f\right)\cdot\bv_{k},
\end{align}
where $\upnot{g}{dl}_{k}\left(t,f\right)$ is the DL channel gain of UT $k$ given by
\begin{align}\label{eq:sat_cha_mod_ray_re_gain}
  \upnot{g}{dl}_{k}\left(t,f\right)
  &\triangleq
\sum_{p=0}^{P_{k}-1}\upnot{g}{dl}_{k,p}\cdot\expx{\barjmath2\pi\left[t\left(\nu_{k,p}-\upnot{\nu}{sat}_{k}\right)
-f\left(\tau_{k,p}-\upnot{\tau}{min}_{k}\right)\right]} \ntb
&=\sum_{p=0}^{P_{k}-1}\upnot{g}{dl}_{k,p}\cdot\expx{\barjmath2\pi\left[t\upnot{\nu}{ut}_{k,p}
-f\upnot{\tau}{ut}_{k,p}\right]},
\end{align}
which will be convenient for derivation of the transmission signal model later.

\subsubsection{Gain}

Note that the statistical properties of the fluctuations of the channel gain $\upnot{g}{dl}_{k}\left(t,f\right)$ in LEO satellite communications mainly depend on the propagation environment in which the UT is located. Note that LEO satellite communication systems are usually operated under line-of-sight (LOS) propagations and Rician channel model is widely accepted in LOS satellite communication systems. In this work, we focus on the case where both non-shadowed LOS and non-LOS paths of the LEO satellite channels exist \cite{Letzepis08Capacity}. Then, the channel gain $\upnot{g}{dl}_{k}\left(t,f\right)$ exhibits the Rician fading distribution with the Rician factor $\kappa_{k}$ and power $\expect{\sqrabs{\upnot{g}{dl}_{k}\left(t,f\right)}}=\gamma_{k}$. In other words, the real and imaginary parts of $\upnot{g}{dl}_{k}\left(t,f\right)$ are independently and identically real-valued Gaussian distributed
with mean $\sqrt{\frac{\kappa_{k}\gamma_{k}}{2\left(\kappa_{k}+1\right)}}$ and variance $\frac{\gamma_{k}}{2\left(\kappa_{k}+1\right)}$, respectively.

\subsection{UL Channel Model}

Using the DL channel modeling approach presented in the above subsections, we briefly investigate the UL channel model for LEO satellite communications in this subsection. Note that the UL channel response is the transpose of the DL channel response in TDD systems, and similar channel model can be obtained. Meanwhile, for FDD systems where the relative carrier frequency difference is small, the physical channel parameters, $P_{k}$, $\nu_{u,p}$, $\tau_{k,p}$, $\vartheta_{k}^\mathrm{x}$, and $\vartheta_{k}^\mathrm{y}$ are almost identical between the UL and DL \cite{Xu04generalized,Ugurlu16multipath,Barriac04Space}.
Thus, the major difference between the UL and DL channels lies in the fast fading path gain terms. Similarly as \eqref{eq:sat_cha_mod_ray_re}, the UL space domain channel response between UT $k$ and the LEO satellite at time $t$ and frequency $f$ can be modeled as
\begin{align}\label{eq:con_wb_cha_mod_rg_del_ul_s}
\upnot{\bg}{ul}_{k}\left(t,f\right)=\expx{\barjmath2\pi\left[t\upnot{\nu}{sat}_{k}-f\upnot{\tau}{min}_{k}\right]}
  \cdot\upnot{g}{ul}_{k}\left(t,f\right)\cdot\bu_{k}\in\bbC^{M\times 1},
\end{align}
where $\upnot{g}{ul}_{k}\left(t,f\right)$ is the UL channel gain of UT $k$ given by
\begin{align}\label{eq:sat_cha_mod_ray_re_gain_ul}
  \upnot{g}{ul}_{k}\left(t,f\right)\triangleq\sum_{p=0}^{P_{k}-1}\upnot{g}{ul}_{k,p}\cdot\expx{\barjmath2\pi\left[t\upnot{\nu}{ut}_{k,p}
-f\upnot{\tau}{ut}_{k,p}\right]},
\end{align}
which exhibits the same statistical properties as the DL channel gain $\upnot{g}{dl}_{k}\left(t,f\right)$, i.e., the real and imaginary parts of $\upnot{g}{ul}_{k}\left(t,f\right)$ are independently and identically real-valued Gaussian distributed
with mean $\sqrt{\frac{\kappa_{k}\gamma_{k}}{2\left(\kappa_{k}+1\right)}}$ and variance $\frac{\gamma_{k}}{2\left(\kappa_{k}+1\right)}$, respectively,
and
$\bu_{k}$ is the UL channel direction vector given by
\begin{align}\label{eq:arr_resp_sing_ul}
\bu_{k}&=\bu_{k}^\mathrm{x}\otimes\bu_{k}^\mathrm{y} \ntb
&=\bu_{\mathrm{x}}\left(\vartheta_{k}^\mathrm{x}\right)\otimes
\bu_{\mathrm{y}}\left(\vartheta_{k}^\mathrm{y}\right)
\in\bbC^{M\times 1},
\end{align}
which exhibits a similar structure as the DL channel direction vector $\bv_{k}$ in \eqref{eq:arr_resp_sing} but with a center frequency offset for FDD systems, and can be well approximated by $\bv_{k}$ when the frequency separation between the UL and the DL is not significant \cite{Barzegar19FDD}.
Similarly as the DL case, the LEO satellite UL channel also exhibits the asymptotic orthogonality as
\begin{align}\label{eq:asy_ort_ul}
  \lim_{M_{d}\to\infty}\left(\bu_{k}^{d}\right)^{H}\bu_{k'}^{d}=\delfunc{k-k'}.
\end{align}
Note that the channel models in \eqref{eq:sat_cha_mod_ray_re} and \eqref{eq:con_wb_cha_mod_rg_del_ul_s} are general in the sense that they take into account the LEO satellite channel propagation properties in the space, time, and frequency domains.

\subsection{DL/UL Transmission Signal Model}

Consider a wideband massive MIMO LEO satellite communication system employing orthogonal frequency division
multiplexing (OFDM) modulation \cite{Papathanassiou01comparison} with the number of subcarriers, $\dnnot{N}{us}$, and the cyclic prefix (CP), $\dnnot{N}{cp}$ samples. Denote by $\dnnot{T}{s}$ the system sampling interval. Then, the OFDM symbol length and the CP length are given by $\dnnot{T}{us}=\dnnot{N}{us}\dnnot{T}{s}$ and $\dnnot{T}{cp}=\dnnot{N}{cp}\dnnot{T}{s}$, respectively. Note that with the delay and Doppler properties of the LEO satellite channels taken into account, it is not difficult to select proper OFDM parameters such that the effects of the intersymbol and intercarrier interference can be almost neglected \cite{Morelli07Synchronization}.

Let $\left\{\upnot{\bx}{dl}_{\ell,n}\right\}_{n=0}^{\dnnot{N}{us}-1}$ be the DL transmit symbols during symbol $\ell$. Then, the transmitted signal $\upnot{\bx_{\ell}}{dl}\parenth{t}\in\bbC^{M\times1}$ can be written as \cite{Hwang09OFDM}
\begin{align}\label{eq:bm_ofdm_tx_n}
\upnot{\bx_{\ell}}{dl}\parenth{t}
=\sum_{n=0}^{\dnnot{N}{us}-1}\upnot{\bx}{dl}_{\ell,n}\cdot\expx{\barjmath2\pi\frac{n}{\dnnot{T}{us}}t},\ -\dnnot{T}{cp}\leq t-\ell\left(\dnnot{T}{cp}+\dnnot{T}{us}\right)<\dnnot{T}{us},
\end{align}
and the corresponding received signal at UT $k$ is given by (where the noise is omitted for brevity)
\begin{align}\label{eq:rec_bm_dl_n}
\upnot{y_{k,\ell}}{dl}\parenth{t}
  &=\int\limits_{-\infty}^{\infty}\!\left[\upnot{\bg_{k}}{dl}\parenth{t,\tau}\right]^{T}\cdot\upnot{\bx_{\ell}}{dl}\parenth{t-\tau}\intdx{\tau},
\end{align}
where $\upnot{\bg_{k}}{dl}\parenth{t,\tau}$ is the inverse Fourier transform of $\upnot{\bg_{k}}{dl}\parenth{t,f}$ in \eqref{eq:sat_cha_mod_ray_re} in terms of $\tau$.

Utilizing the Doppler and delay properties of the LEO satellite propagation channels addressed previously, we proceed to perform time and frequency synchronization. In particular, with delay compensation $\upnot{\tau}{syn}_{k}=\upnot{\tau}{min}_{k}$ and Doppler compensation $\upnot{\nu}{syn}_{k}=\upnot{\nu}{sat}_{k,p}$
applied to the received signal at UT $k$, the resultant signal can be represented by
\begin{align}\label{eq:mul_sync_n}
\upnot{y_{k,\ell}}{dl,syn}\parenth{t}
=\upnot{y_{k,\ell}}{dl}\parenth{t+\upnot{\tau}{syn}_{k}}\cdot\expx{-\barjmath2\pi\left(t+\upnot{\tau}{syn}_{k}\right)\upnot{\nu}{syn}_{k}}.
\end{align}
Then, the corresponding signal dispersion in the delay and Doppler domains can be significantly reduced, and it is not difficult to select proper OFDM parameters to mitigate the intersymbol and intercarrier interference \cite{Hwang09OFDM}. Consequently, the demodulated DL received signal at UT $k$ over subcarrier $n$ of OFDM symbol $\ell$ can be represented by
\begin{align}\label{eq:sat_dl_re}
  \upnot{y}{dl}_{k,\ell,n}=\left(\upnot{\bg}{dl}_{k,\ell,n}\right)^{T}\upnot{\bx}{dl}_{\ell,n},
\end{align}
where $\upnot{\bg}{dl}_{k,\ell,n}$ is the DL channel of UT $k$ over symbol $\ell$ and subcarrier $n$ given by \cite{Hwang09OFDM}
\begin{align}\label{eq:sat_dl_dis_ch}
  \upnot{\bg}{dl}_{k,\ell,n}
  =\bv_{k}\cdot\upnot{g}{dl}_{k,\ell,n}\in\bbC^{M\times 1},
\end{align}
where $\upnot{g}{dl}_{k,\ell,n}
  =\upnot{g}{dl}_{k}\left(\ell\left(\dnnot{T}{us}+\dnnot{T}{cp}\right),n/\dnnot{T}{us}\right)$.

Besides, consider UL transmission employing OFDM modulation with similar parameters as DL transmission. Then, with proper delay and Doppler compensations performed at the UT side, the demodulated UL received signal at the satellite over symbol $\ell$ and subcarrier $n$ can be represented as
\begin{align}\label{eq:sat_dl_re_ul}
  \upnot{\by}{ul}_{\ell,n}=\sum_{k}\upnot{\bg}{ul}_{k,\ell,n}\upnot{x}{ul}_{k,\ell,n}\in\bbC^{M\times 1},
\end{align}
where $\upnot{x}{ul}_{k,\ell,n}$ is the complex-valued symbols transmitted by UT $k$, and $\upnot{\bg}{ul}_{k,\ell,n}$ is the UL channel of UT $k$ over subcarrier $n$ of OFDM symbol $\ell$ given by
\begin{align}\label{eq:sat_dl_dis_ch_ul}
  \upnot{\bg}{ul}_{k,\ell,n}
  =\bu_{k}\cdot\upnot{g}{ul}_{k,\ell,n}\in\bbC^{M\times 1},
\end{align}
where $\upnot{g}{ul}_{k,\ell,n}
  =\upnot{g}{ul}_{k}\left(\ell\left(\dnnot{T}{us}+\dnnot{T}{cp}\right),n/\dnnot{T}{us}\right)$.
Note that the transmission signal models in \eqref{eq:sat_dl_re} and \eqref{eq:sat_dl_re_ul} are applicable provided that delay and Doppler compensations are properly performed via exploiting the delay and Doppler properties of the LEO satellite channels described previously.

\section{Statistical CSI based DL/UL Transmissions}\label{sec:ud_trans}

In this section, we investigate DL precoder and UL receiver design for LEO satellite communications based on the channel and signal models established in the above section. Note that the conventional designs of DL precoding vectors and UL receiving vectors in MIMO transmission usually require knowledge of iCSI. However, it is in general infeasible to obtain precise iCSI at the satellite sides for DL of LEO satellite communications. In addition, frequent update of the DL precoding vectors and UL receiving vectors using iCSI will be challenging for implementation on payload of practical satellite communications. Hereafter, we focus on the design of DL precoder and UL receiver utilizing slowly-varying sCSI for satellite communications.

\subsection{DL Precoder}

We first consider DL transmission where $K$ single antenna UTs are simultaneously served in the same time-frequency blocks, and the served UT set is denoted by $\cK=\left\{0,1,\ldots,K-1\right\}$. For DL linear precoding performed at the satellite, the signal received by UT $k\in\cK$ in \eqref{eq:sat_dl_re} can be rewritten as
\begin{align}\label{eq:sat_dl_re2}
  \upnot{y}{dl}_{k}
  =\left(\upnot{\bg}{dl}_{k}\right)^{T}
  \sum_{i\in\cK}\sqrt{\upnot{q}{dl}_{i}}\bb_{i}\upnot{s}{dl}_{i}+\upnot{z}{dl}_{k},
\end{align}
where the subcarrier and symbol indices are omitted for brevity, $\upnot{q}{dl}_{k}$ is the transmit power allocated to UT $k$, $\bb_{k}\in\bbC^{M\times1}$ is the normalized transmit precoding vector satisfying the $\normt{\bb_{k}}=\sqrt{\bb_{k}^{H}\bb_{k}}=1$, $\upnot{s}{dl}_{k}$ is the signal for UT $k$ with mean $0$ and variance $1$, and $\upnot{z}{dl}_{k}$ is the additive circular symmetric complex-valued Gaussian noise with mean $0$ and variance $\upnot{\sigma}{dl}_{k}$, i.e., $\upnot{z}{dl}_{k}\sim\GCN{0}{\upnot{\sigma}{dl}_{k}}$.

Note that SLNR is a convenient and efficient design metric widely adopted in DL multiuser MIMO transmission, and we first review the SLNR maximization criterion based precoding approach. In particular, the SLNR of UT $k$ in the DL is given by \cite{Sadek07leakage,Patcharamaneepakorn12On}
\begin{align}\label{eq:dl_slnr_k}
  \mathsf{SLNR}_{k}
  =\frac{\sqrabs{\left(\upnot{\bg}{dl}_{k}\right)^{T}\bb_{k}}\upnot{q}{dl}_{k}}
  {\sum_{i\neq k}\sqrabs{\left(\upnot{\bg}{dl}_{i}\right)^{T}\bb_{k}}\upnot{q}{dl}_{k}+\upnot{\sigma}{dl}_{k}}
  =\frac{\sqrabs{\left(\upnot{\bg}{dl}_{k}\right)^{T}\bb_{k}}}
  {\sum_{i\neq k}\sqrabs{\left(\upnot{\bg}{dl}_{i}\right)^{T}\bb_{k}}+\oneon{\upnot{\rho}{dl}_{k}}},
\end{align}
where $\upnot{\rho}{dl}_{k}\triangleq\upnot{q}{dl}_{k}/\upnot{\sigma}{dl}_{k}$ is the DL signal-to-noise ratio (SNR) of UT $k$. Then the precoder of UT $k$ that maximizes $\mathsf{SLNR}_{k}$ in \eqref{eq:dl_slnr_k} can be obtained as
\begin{align}\label{eq:dl_slnr_k_pre}
  \upnot{\bb}{slnr}_{k}
  =\oneon{\upnot{\eta}{slnr}_{k}}\left[\left(\sum_{i}\upnot{\bg}{dl}_{i}\left(\upnot{\bg}{dl}_{i}\right)^{H}
  +\oneon{\upnot{\rho}{dl}_{k}}\bI_{M}\right)^{-1}\upnot{\bg}{dl}_{k}\right]^{*},
\end{align}
where $(\cdot)^{*}$ denotes the conjugate operation and $\upnot{\eta}{slnr}_{k}$ is the power normalization coefficient that is set to satisfy $\normt{\upnot{\bb}{slnr}_{k}}=1$.
We mention that the SLNR maximization DL precoder in \eqref{eq:dl_slnr_k_pre} requires knowledge of iCSI $\upnot{\bg}{dl}_{k}$ for all $k$. However, it is in general difficult to obtain precise DL iCSI for transmitter at the satellite side.

In the following, we investigate DL precoding for satellite communications using long-term sCSI at the transmitter, including the channel direction vector $\bv_{k}$ and the statistics of the channel gain $\upnot{g}{dl}_{k,\ell,n}$.
We consider the ASLNR performance metric as follows \cite{Joung09Relay}
\begin{align}\label{eq:dl_slnr_k_pre_lb}
  \mathsf{ASLNR}_{k}&\triangleq
  \frac{\expect{\sqrabs{\left(\upnot{\bg}{dl}_{k}\right)^{T}\bb_{k}}}}
  {\expect{\sum_{i\neq k}\sqrabs{\left(\upnot{\bg}{dl}_{i}\right)^{T}\bb_{k}}+\oneon{\upnot{\rho}{dl}_{k}}}}
  =\frac{\gamma_{k}\sqrabs{\left(\bv_{k}\right)^{T}\bb_{k}}}
  {\sum_{i\neq k}\gamma_{i}\sqrabs{\left(\bv_{i}\right)^{T}\bb_{k}}+\oneon{\upnot{\rho}{dl}_{k}}}
  ,
\end{align}
where the numerator and the denominator account for the average power of the signal and leakage plus noise, respectively.
The sCSI based precoder that maximizes $\mathsf{ASLNR}_{k}$ is presented in the following proposition.
\begin{prop}\label{prop:dl_pre}
The precoding vector that maximizes $\mathsf{ASLNR}_{k}$ in \eqref{eq:dl_slnr_k_pre_lb} is given by
\begin{align}\label{eq:dl_aslnr_k_pre_lb}
  \upnot{\bb}{aslnr}_{k}
  =\oneon{\upnot{\eta}{aslnr}_{k}}\left[\left(\sum_{i}\gamma_{i}\bv_{i}\bv_{i}^{H}
  +\oneon{\upnot{\rho}{dl}_{k}}\bI_{M}\right)^{-1}\bv_{k}\right]^{*},
\end{align}
where $\upnot{\eta}{aslnr}_{k}$ is the power normalization coefficient that is set to satisfy $\normt{\upnot{\bb}{aslnr}_{k}}=1$,
and the corresponding maximum ASLNR value is given by
\begin{align}\label{eq:dl_aslnr_k_max}
  \upnot{\mathsf{ASLNR}}{max}_{k}
  =\frac{1}{1-\gamma_{k}\bv_{k}^{H}\left(\sum_{i}\gamma_{i}\bv_{i}\bv_{i}^{H}
  +\oneon{\upnot{\rho}{dl}_{k}}\bI_{M}\right)^{-1}\bv_{k}}-1.
\end{align}
\end{prop}
\begin{IEEEproof}
The proof is similar to the iCSI case in \cite{Patcharamaneepakorn12On}, and is omitted for brevity.
\end{IEEEproof}

\propref{prop:dl_pre} provides a sCSI based DL precoder that maximizes the ASLNR in closed-form. Note that the sCSI required in the proposed approach are the channel direction vectors and the average power of all UTs' channels, i.e., $\bv_{k}$ and $\gamma_{k}$, $\forall k$.
In addition, from the definition of the channel direction vector in \eqref{eq:arr_resp_sing}, only the space angles, $\vartheta_{k}^\mathrm{x}$ and $\vartheta_{k}^\mathrm{y}$, are needed for estimating $\bv_{k}$. Thus, the number of parameters in statistical CSI to estimate can be significantly reduced. As the proposed sCSI based DL precoding design is independent of subcarriers and OFDM symbols in transmission interval where the channels statistics do not change significantly and thus is convenient for practical implementation of the satellite payloads.
Then, the computational overhead for DL precoding design can be reduced compared with the iCSI based approach.

\subsection{UL Receiver}

In this subsection, we investigate UL receiver design.
The UL received signal by the satellite in \eqref{eq:sat_dl_re_ul} can be rewritten as
\begin{align}\label{eq:sat_dl_re_ul_re}
  \upnot{\by}{ul}=\sum_{k}\upnot{\bg}{ul}_{k}\sqrt{\upnot{q}{ul}}\upnot{s}{ul}_{k}+\upnot{\bz}{ul},
\end{align}
where the subcarrier and symbol indices are omitted for brevity, $\upnot{q}{ul}$ is the transmit power of one UT, $\upnot{s}{ul}_{k}$ is the signal sent by UT $k$ with mean $0$ and variance $1$, and $\upnot{\bz}{ul}$ is the additive Gaussian noise distributed as $\GCN{\bzero}{\upnot{\sigma}{ul}\bI_{M}}$.
With a linear receiver at the satellite, the recovered signal of UT $k$ can be expressed by
\begin{align}\label{eq:sat_ul_rec}
  \upnot{\hat{s}}{ul}_{k}
  =\bw_{k}^{T}\upnot{\by}{ul}
  =\bw_{k}^{T}\sum_{i}\upnot{\bg}{ul}_{i}\sqrt{\upnot{q}{ul}}\upnot{s}{ul}_{i}+\bw_{k}^{T}\upnot{\bz}{ul},
\end{align}
where $\bw_{k}$ is the linear receiving vector of UT $k$. Then the SINR of UT $k$ is given by
\begin{align}\label{eq:sat_ul_sinr}
  \mathsf{SINR}_{k}
  =\frac{\sqrabs{\bw_{k}^{T}\upnot{\bg}{ul}_{k}}\upnot{q}{ul}}
  {\sum_{i\neq k}\sqrabs{\bw_{k}^{T}\upnot{\bg}{ul}_{i}}\upnot{q}{ul}
  +\upnot{\sigma}{ul}\sqrnormt{\bw_{k}}}
  =\frac{\sqrabs{\bw_{k}^{T}\upnot{\bg}{ul}_{k}}}
  {\sum_{i\neq k}\sqrabs{\bw_{k}^{T}\upnot{\bg}{ul}_{i}}
  +\oneon{\upnot{\rho}{ul}}\sqrnormt{\bw_{k}}},
\end{align}
where $\upnot{\rho}{ul}\triangleq\upnot{q}{ul}/\upnot{\sigma}{ul}$ is the UL SNR.
It is not difficult to obtain the receiver of UT $k$ that maximizes $\mathsf{SINR}_{k}$ in \eqref{eq:sat_ul_sinr} as
\begin{align}\label{eq:ul_sinr_k_rec}
  \upnot{\bw}{sinr}_{k}
  =\left[\left(\sum_{i}\upnot{\bg}{ul}_{i}\left(\upnot{\bg}{ul}_{i}\right)^{H}
  +\oneon{\upnot{\rho}{ul}}\bI_{M}\right)^{-1}\upnot{\bg}{ul}_{k}\right]^{*}.
\end{align}
Note that the SINR maximization UL receiving vectors in \eqref{eq:ul_sinr_k_rec} are in general difficult to be computed in practical satellite communications systems where the payload resource is limited, as they are needed to be updated more frequently in time and frequency.

Similarly as the DL case, we investigate UL receiver design for LEO satellite communications exploiting sCSI and consider the ASINR performance metric given by
\begin{align}\label{eq:ul_sinr_k_rec_lb}
  \mathsf{ASINR}_{k}&\triangleq
  \frac{\expect{\sqrabs{\bw_{k}^{T}\upnot{\bg}{ul}_{k}}}}
  {\expect{\sum_{i\neq k}\sqrabs{\bw_{k}^{T}\upnot{\bg}{ul}_{i}}
  +\oneon{\upnot{\rho}{ul}}\sqrnormt{\bw_{k}}}}
  =\frac{\gamma_{k}\sqrabs{\left(\bu_{k}\right)^{T}\bw_{k}}}
  {\sum_{i\neq k}\gamma_{i}\sqrabs{\left(\bu_{i}\right)^{T}\bw_{k}}+\oneon{\upnot{\rho}{ul}}\sqrnormt{\bw_{k}}}
  .
\end{align}
Using a similar proof procedure as in \propref{prop:dl_pre}, we can obtain that the sCSI based UL receiver that maximizes $\mathsf{ASINR}_{k}$ in \eqref{eq:ul_sinr_k_rec_lb} is given by
\begin{align}\label{eq:ul_asinr_k_rec_lb}
  \upnot{\bw}{asinr}_{k}
  =\left[\left(\sum_{i}\gamma_{i}\bu_{i}\bu_{i}^{H}
  +\oneon{\upnot{\rho}{ul}}\bI_{M}\right)^{-1}\bu_{k}\right]^{*},
\end{align}
with the corresponding maximum ASINR of UT $k$ being
\begin{align}\label{eq:ul_asinr_k_max}
  \upnot{\mathsf{ASINR}}{max}_{k}
  =\frac{1}{1-\gamma_{k}\bu_{k}^{H}\left(\sum_{i}\gamma_{i}\bu_{i}\bu_{i}^{H}
  +\oneon{\upnot{\rho}{ul}}\bI_{M}\right)^{-1}\bu_{k}}-1.
\end{align}
Note that the sCSI based UL receiver in \eqref{eq:ul_asinr_k_rec_lb} is presented in closed-form, and is based on sCSI, i.e., the channel direction vector $\bu_{k}$ and the statistics of the channel gain $\upnot{g}{dl}_{k,\ell,n}$, which can thus mitigate the payload complexity and cost in practical satellite communications.

\subsection{DL-UL Duality}

From \eqref{eq:dl_aslnr_k_pre_lb} and \eqref{eq:ul_asinr_k_rec_lb}, we can obtain the DL-UL duality between the proposed sCSI based DL precoder and UL receiver. Specifically, in the considered transmission interval where the channel statistics do not change significantly, if the DL data transmission SNR $\upnot{\rho}{dl}_{k}$ equals the UL data transmission SNR $\upnot{\rho}{ul}$, then the sCSI based DL precoding vectors in \eqref{eq:dl_aslnr_k_pre_lb} are equal to the sCSI based UL receiving vectors in \eqref{eq:ul_asinr_k_rec_lb} with proper power normalization provided that the DL direction vector $\bv_{k}$ equals the UL direction vector $\bu_{k}$, and the transmission complexity can be further reduced. Note that different from the UL-DL duality results based on the perfect iCSI assumption in, e.g., \cite{Viswanath03Sum} and \cite{Shi07Downlink}, our result is established using the sCSI at the satellite side.

\subsection{Upper Bound of ASLNR/ASINR}

In this subsection, we investigate the conditions under which the DL ASLNR and UL ASINR metrics considered above can be upper bounded.

\begin{prop}\label{prop:dl_op}
The maximum DL ASLNR value $\upnot{\mathsf{ASLNR}}{max}_{k}$ in \eqref{eq:dl_aslnr_k_max} is upper bounded by
\begin{align}\label{eq:dl_aslnr_ub}
  \upnot{\mathsf{ASLNR}}{max}_{k}
  \leq\upnot{\rho}{dl}_{k}\gamma_{k},
\end{align}
and the upper bound can be achieved under the condition that
\begin{align}\label{eq:dl_aslnr_k_max_con_ac}
  \left(\bv_{k}^\mathrm{x}\right)^{H}\bv_{i}^\mathrm{x}=0 \quad\mathrm{or}\quad
  \left(\bv_{k}^\mathrm{y}\right)^{H}\bv_{i}^\mathrm{y}=0 ,\quad\forall k\neq i.
\end{align}
Besides, the maximum UL ASINR value $\upnot{\mathsf{ASINR}}{max}_{k}$ in \eqref{eq:ul_asinr_k_max} is upper bounded by
\begin{align}\label{eq:ul_asinr_ub}
  \upnot{\mathsf{ASINR}}{max}_{k}
  \leq\upnot{\rho}{ul}\gamma_{k},
\end{align}
and the upper bound can be achieved under the condition that
\begin{align}\label{eq:dl_asinr_k_max_con_ac}
  \left(\bu_{k}^\mathrm{x}\right)^{H}\bu_{i}^\mathrm{x}=0 \quad\mathrm{or}\quad
  \left(\bu_{k}^\mathrm{y}\right)^{H}\bu_{i}^\mathrm{y}=0 ,\quad\forall k\neq i.
\end{align}
\end{prop}
\begin{IEEEproof}
Please refer to \appref{app:prop_dl_op}.
\end{IEEEproof}

\propref{prop:dl_op} shows that the DL ASLNRs and UL ASINRs of all served UTs with the proposed sCSI based precoder and receiver can reach their upper bounds provided that the corresponding channel direction vectors of different UTs are mutually orthogonal. The result in \propref{prop:dl_op} is physically intuitive as the DL channel leakage power and the UL inter-user interference can be eliminated provided that the conditions in \eqref{eq:dl_aslnr_k_max_con_ac} and \eqref{eq:dl_asinr_k_max_con_ac} are satisfied.

From \eqref{eq:asy_ort_dl} and \eqref{eq:asy_ort_ul}, we can observe that the optimal conditions obtained in \propref{prop:dl_op} can be asymptotically satisfied when the number of antennas $M$ tends to infinity. This corroborates the rationality and potential of exploiting massive MIMO in enhancing the transmission performance of satellite communications.

\begin{remark}
When the channel direction vectors of the UTs scheduled over the same time-frequency resource blocks satisfy the conditions in \eqref{eq:dl_aslnr_k_max_con_ac} and \eqref{eq:dl_asinr_k_max_con_ac} or the number of antennas at the satellite side is sufficiently large, we can obtain from the matrix inversion lemma that the proposed sCSI based precoder/receiver in \eqref{eq:dl_aslnr_k_pre_lb} and \eqref{eq:ul_asinr_k_rec_lb} will reduce to
\begin{align}\label{eq:dl_mf}
  \upnot{\bb}{aslnr}_{k}
  =\bv_{k}^{*},\quad
  \upnot{\bw}{asinr}_{k}
  =\bu_{k}^{*}.
\end{align}
Notably, the sCSI based DL precoder and UL receiver presented in \eqref{eq:dl_mf} approach the ones using iCSI as the number of antennas tends to infinity \cite{Marzetta10Noncooperative}, which demonstrates the asymptotic optimality of the proposed precoder/receiver exploiting sCSI.
\end{remark}

\begin{remark}
Note that for the case with a sufficiently large number of antennas at the satellite side, the precoder/receiver in \eqref{eq:dl_mf} will asymptotically tend to the discrete Fourier transform (DFT) based fixed precoder/receiver as follows
\begin{align}\label{eq:dl_dft}
  \bb_{k}=\left[\bv_{\mathrm{x}}\left(\overline{\vartheta}_{k}^\mathrm{x}\right)\otimes
\bv_{\mathrm{y}}\left(\overline{\vartheta}_{k}^\mathrm{y}\right)\right]^{*},\quad
  \upnot{\bw}{asinr}_{k}
  =\bw_{k}=\left[\bu_{\mathrm{x}}\left(\overline{\vartheta}_{k}^\mathrm{x}\right)\otimes
\bu_{\mathrm{y}}\left(\overline{\vartheta}_{k}^\mathrm{y}\right)\right]^{*},
\end{align}
where $\overline{\vartheta}_{k}^{d}$ is the nearest point of $\vartheta_{k}^{d}$ in the DFT grid satisfying $\overline{\vartheta}_{k}^{d}=-1+2n_{k}^{d}/M_{d}$ with $n_{k}^{d}\in\left[0,M_{d}-1\right]$ being integers and $\abs{\overline{\vartheta}_{k}^{d}-\vartheta_{k}^{d}}<2/M_{d}$ for $d\in\mathcal{D}$.  In this case, the precoding/receiving vectors for the simultaneously served UTs in the same user group are orthogonal, and can be efficiently implemented with fast Fourier transform (FFT).
\end{remark}

\section{User Grouping}\label{sec:ut_sc}

From the results in the above section, we can observe that the performance of the proposed sCSI based precoder and receiver in massive MIMO LEO satellite communications will largely depend on the channel statistics of the simultaneously served UTs. As the number of the UTs to be served is usually much larger than that of antennas equipped at the satellites, user grouping is of practical importance. Compared with the terrestrial counterpart, user grouping is of greater interest as the satellite service provider generally aims at serving all UTs
in satellite communications. In this section, we investigate user grouping for massive MIMO LEO satellite communications.

\subsection{Space Angle based User Grouping}\label{subsec:aoduts}

Although the conditions in \propref{prop:dl_op} are desirable for optimizing the performance of DL ASLNRs and UL ASINRs in satellite communications, it is in general difficult to schedule the UTs that rigorously satisfy this condition, and the optimal user grouping pattern can be found through exhaustive search. However, due to the large number of existing UTs in satellite communications, it is usually infeasible to perform an exhaustive search in practical systems.

The optimal user grouping condition presented in \propref{prop:dl_op} indicates that the channel direction vectors of UTs in the same group should be as orthogonal as possible. From the definitions in  \eqref{eq:arr_resp_sing} and \eqref{eq:arr_resp_sing_ul}, the channel direction vectors are directly related to the channel propagation properties in the space domain, i.e., the channel space angles. Then, the conditions for achieving the upper bounds of ASLNR and ASINR presented in \eqref{eq:dl_aslnr_k_max_con_ac} and \eqref{eq:dl_asinr_k_max_con_ac} can be reduced to the condition that the channel space angles should satisfy
\begin{align}\label{eq:dl_aslnr_k_max_con_ac_ang}
  \vartheta_{k}^\mathrm{x}-\vartheta_{i}^\mathrm{x}=\frac{2}{\dnnot{M}{x}}\upnot{n_{k,i}}{x} \quad\mathrm{or}\quad
  \vartheta_{k}^\mathrm{y}-\vartheta_{i}^\mathrm{y}=\frac{2}{\dnnot{M}{y}}\upnot{n_{k,i}}{y} ,\quad\forall k\neq i,
\end{align}
where both $\upnot{n_{k,i}}{x}$ and $\upnot{n_{k,i}}{y}$ are non-zero integers. Motivated by the condition in \eqref{eq:dl_aslnr_k_max_con_ac_ang}, we propose a space angle based user grouping (SAUG) approach as follows. Specifically, we uniformly divide the space angle range $[-1,1)$ into $\dnnot{M}{x} \dnnot{G}{x}$ and $\dnnot{M}{y} \dnnot{G}{y}$ equal sectors in the x- and y-axes, respectively, where $\dnnot{G}{x}$ and $\dnnot{G}{y}$ are both integers and their physical meaning will be clear later.
Then, the space angle intervals after division can be represented by
\begin{align}
\mathcal{A}_{\left( g,r \right)}^{\left(m,n\right)}
&= \left\{ \left. \left( \dnnot{\phi}{x},\dnnot{\phi}{y} \right) \right| \dnnot{\phi}{x} \in \left[  \phi_{g,m}^{\mathrm{x}} - \frac{\Delta_{\mathrm{x}}}{2},\phi_{g,m}^{\mathrm{x}} + \frac{\Delta_{\mathrm{x}}}{2} \right)   ,\
 \dnnot{\phi}{y} \in \left[  \phi_{r,n}^{\mathrm{y}} - \frac{\Delta_{\mathrm{y}}}{2},\phi_{r,n}^{\mathrm{y}} + \frac{\Delta_{\mathrm{y}}}{2} \right)   \right\},
\end{align}
where $\phi_{a,b}^d$ for $d \in \mathcal{D}$ is the center space angle of the interval in the x-/y-axis given by
\begin{align}
\phi_{a,b}^d = -1 + \frac{\Delta_d}{2} + (a + b G_d) \Delta_d ,\quad 0 \le a \le G_d -1,\, 0 \le b \le M_d-1,
\end{align}
with $\Delta_d = 2/\left( M_d G_d \right)$ being the length of the space angle interval in the x-/y-axis.

With the above definition of the space angle interval division, the UTs can be grouped as follows.
A given UT $k$ is scheduled into the $(g,r)$th group if there exist $0 \le m \le M_{\mathrm{x}}-1$ and $0 \le n \le M_{\mathrm{y}}-1$ such that
the corresponding channel space angles satisfy
\begin{align}
\left( \vartheta_k^{\mathrm{x}},\vartheta_k^{\mathrm{y}} \right) \in  \mathcal{A}_{\left(g,r\right)}^{\left(m,n\right)}.
\end{align}
Denote by $\mathcal{K}_{\left( g,r \right)}^{(m,n)} = \left\{ k : \left( \vartheta_k^{\mathrm{x}},\vartheta_k^{\mathrm{y}} \right) \in  \mathcal{A}_{\left(g,r\right)}^{\left(m,n\right)}   \right\}$ the set of UTs whose space angles lie in the interval $\mathcal{A}_{\left(g,r\right)}^{\left(m,n\right)}$. In the proposed SAUG approach, we always require
$\left\lvert \mathcal{K}_{\left( g,r \right)}^{(m,n)} \right\rvert \le 1$ to avoid intra-beam interference. Note that other UTs located in the same space angle interval can be scheduled over different time-frequency resources in a round-robin manner to preserve fairness. Based on the above user grouping procedure, the UTs are scheduled into at most $G_{\mathrm{x}} G_{\mathrm{y}}$ groups, where the $\left(g,r\right)$th UT group is defined as
\begin{align}\label{eq:utgr}
\mathcal{K}_{\left(g,r\right)}\triangleq
\bigcup_{ \substack{0 \le m \le M_{\mathrm{x}}-1 \\ 0 \le n \le M_{\mathrm{y}}-1} }\mathcal{K}_{\left(g,r\right)}^{\left(m,n\right)}.
\end{align}
Note that the UTs scheduled in the same group will perform transmission over the same time-frequency resources, while UTs in different groups will be allocated with different time-frequency transmission resources.

\subsection{Achievable Rate Performance}

In this subsection, we investigate the achievable rate performance of the proposed SAUG approach. We focus on the DL transmission case, and the UL results can be similarly obtained.

From the DL signal model in \eqref{eq:sat_dl_re2} and the proposed SAUG approach in the above subsection, the DL achievable ergodic sum rate is given by
\begin{align}\label{eq:dlralsnr}
\dnnot{R}{dl} = \oneon{G_{\mathrm{x}}G_{\mathrm{y}}} \sum_{g=0}^{G_{\mathrm{x}}-1} \sum_{r=0}^{G_{\mathrm{y}}-1} \sum_{k\in\cK_{\left(g,r\right)}}\expect{ \log_2 \left\{ 1 + \frac{ \left\lvert g_{k}^{\DL} \right\rvert^2 \left\lvert \bv_k^T \bb_k^{\ASLNR} \right\rvert^2 q_k^{\DL} }{ \sum_{i \in\cK_{\left(g,r\right)}}^{i\ne k} \left\lvert g_{k}^{\DL} \right\rvert^2 \left\lvert \bv_k^T \bb_i^{\ASLNR} \right\rvert^2 q_i^{\DL} + \sigma_k^{\DL}  } \right\} },
\end{align}
where $\bb_k^{\ASLNR}$ is the sCSI based precoder of UT $k$ presented in \eqref{eq:dl_aslnr_k_pre_lb}, and $\cK_{\left(g,r\right)}$ is the UT group defined in \eqref{eq:utgr}. The ergodic rate expression in \eqref{eq:dlralsnr} is in general difficult to handle. Therefore, we resort to investigate the bounds of the achievable ergodic rate for further analysis.
In the following proposition, we first present an upper bound of the DL achievable ergodic sum rate.

\begin{prop}\label{prop:sp_rate_max_cond}
With linear precoder utilizing only sCSI, the DL achievable ergodic sum rate $\dnnot{R}{dl}$ in \eqref{eq:dlralsnr} is upper bounded by
\begin{align}\label{eq:upp_rk}
  \dnnot{R}{dl}
  \leq \dnnot{R}{dl}^{\mathrm{ub}}
  &\triangleq \oneon{G_{\mathrm{x}}G_{\mathrm{y}}}\sum_{g=0}^{G_{\mathrm{x}}-1} \sum_{r=0}^{G_{\mathrm{y}}-1} \sum_{k\in\cK_{\left(g,r\right)}}
  \expect{\log_{2}\left\{1+\upnot{\rho}{dl}_{k}\sqrabs{\upnot{g}{dl}_{k}}\right\}},
\end{align}
where the corresponding upper bound can be achieved provided that the channel direction vectors of the UTs served in the same group satisfy
\begin{subequations}
\begin{align}
  \left(\bv_{k}^\mathrm{x}\right)^{H}\bv_{i}^\mathrm{x}=0\quad\mathrm{or}\quad
  \left(\bv_{k}^\mathrm{y}\right)^{H}\bv_{i}^\mathrm{y}=0,&\quad\forall k,i\in\cK_{\left(g,r\right)},\,k\neq i,\label{eq:upp_rk_conda} \\
  \bb_k^{\ASLNR}=\left(\bv_{k}^\mathrm{x}\otimes\bv_{k}^\mathrm{y}\right)^{*},&\quad \forall k. \label{eq:upp_rk_condb}
\end{align}
\end{subequations}
\end{prop}
\begin{IEEEproof}
Please refer to \appref{app:prop_dl_upbound}.
\end{IEEEproof}

\propref{prop:sp_rate_max_cond} provides some insights for optimal DL precoding design with sCSI at the transmitter. In particular, the channel direction vectors of the UTs scheduled to be served over the same time-frequency resources should be as orthogonal as possible. Meanwhile, the beamforming vector of a given UT should be aligned with the corresponding channel direction vector. Note that the previously proposed SAUG approach attempts to schedule the UTs to satisfy the condition in \eqref{eq:upp_rk_conda}, and the proposed ASLNR based precoding strives to reduce the inter-user interference to approach the condition in \eqref{eq:upp_rk_condb} as remarked in \eqref{eq:dl_mf}.

In the following, we further investigate the asymptotic performance of the proposed approach.
Before proceeding, we first provide an upper bound of the inner product $\left\lvert \bv_k^H \bv_j \right\rvert$ for UTs $k,j(\forall k\neq j)$ that are scheduled over the same time-frequency transmission resources via the proposed SAUG approach.
From \eqref{eq:ula steer dm}, we have
\begin{equation}
\left\lvert \bv_k^H \bv_j \right\rvert = \left\lvert \frac{ \sin  \frac{\pi \varphi_{\mathrm{x}} M_{\mathrm{x}}}{2}  }{ M_{\mathrm{x}} \sin  \frac{\pi \varphi_{\mathrm{x}}}{2}  } \right\rvert  \left\lvert \frac{ \sin  \frac{\pi \varphi_{\mathrm{y}} M_{\mathrm{y}}}{2}  }{ M_{\mathrm{y}} \sin  \frac{\pi \varphi_{\mathrm{y}}}{2} }  \right\rvert,\quad
\forall k \neq j. \label{inner_product_vkvj_sinc}
\end{equation}
where $\varphi_{d} = \vartheta_{k}^{d} - \vartheta_{j}^{d}$ for $d \in \mathcal{D}$.
With the proposed SAUG approach, it is not difficult to show that
\begin{equation}
\frac{2}{M_d} m - \Delta_d \le \varphi_d \le \frac{2}{M_d} m + \Delta_d, \text{ where } 1 \le m \le M_d - 1, d \in \mathcal{D}.
\end{equation}
Then, we can further upper bound the inner product $\left\lvert \bv_k^H \bv_j \right\rvert$ as
\begin{equation}
\left\lvert \bv_k^H \bv_j \right\rvert \le \left\lvert  \frac{ \sin \pi \left( 1 - \frac{1}{G_{\mathrm{x}}} \right) }{ M_{\mathrm{x}} \sin \frac{\pi}{M_{\mathrm{x}}} \left( 1 - \frac{1}{G_{\mathrm{x}}} \right)  } \right\rvert \left\lvert  \frac{ \sin \pi \left( 1 - \frac{1}{G_{\mathrm{y}}} \right) }{ M_{\mathrm{y}} \sin \frac{\pi}{M_{\mathrm{y}}} \left( 1 - \frac{1}{G_{\mathrm{y}}} \right)  } \right\rvert,\quad \forall k \ne j.
\end{equation}
Thus, with sufficiently large numbers of groups $G_{\mathrm{x}}$ and $G_{\mathrm{y}}$, the inner product $\left\lvert \bv_k^H \bv_j \right\rvert$ can be sufficiently small. Motivated by this, we present the asymptotic optimality of the proposed SAUG approach combined with sCSI based DL precoder in each UT group in the following proposition.

\begin{prop} \label{prop:dl_lowerbound}
The DL achievable ergodic sum rate $\dnnot{R}{dl}$ with the proposed sCSI based ASLNR maximization DL precoder and the SAUG approach is lower bounded by
	\begin{equation}\label{eq:prop_lb}
	R_{\DL}\geq\upnot{R}{lb}_{\DL} \triangleq
\oneon{G_{\mathrm{x}}G_{\mathrm{y}}}\sum_{g=0}^{G_{\mathrm{x}}-1} \sum_{r=0}^{G_{\mathrm{y}}-1} \sum_{k\in\cK_{\left(g,r\right)}}
\expect{ \log_2 \left\{ 1 + \frac{ \left\lvert g_{k}^{\DL} \right\rvert^2 \left( 1 - \delta^{\DL} \left( \epsilon \right) \right) q_k^{\DL} }{ \sum^{i \ne k}_{i\in\cK_{\left(g,r\right)}} \left\lvert g_{k}^{\DL} \right\rvert^2 \frac{  \beta_{k,i}^{\DL}(\epsilon) }{ \xi^{\DL}(\epsilon)  } q_i^{\DL} + \sigma_k^{\DL}  } \right\} },
	\end{equation}
	where $\delta^{\DL} \left( \epsilon \right)$, $\beta_{k,i}^{\DL}(\epsilon)$, and $\xi^{\DL}(\epsilon)$ are given by
	\begin{subequations}
		\begin{align}
		\delta^{\DL} \left( \epsilon \right) &= \frac{ \left( \rho_{\MAX}^{\DL}\gamma_{\MAX} \right)^2 (K_{\MAX}-1)^2 \epsilon^2 }{ \chi^{\DL}(\epsilon)  }, \\
		\chi^{\DL}(\epsilon) &= \frac{1}{ \frac{1}{\rho_{\MIN}^{\DL} \gamma_{\MIN} } + 1 + (K_{\MAX}-1) \epsilon }, \\
		\beta_{k,i}^{\DL} \left( \epsilon \right) &= \frac{\left( \rho_{\MAX}^{\DL}\gamma_{\MAX} \right)^4}{ \left( \rho_{k}^{\DL} \gamma_i \gamma_k \right)^2 }  (K_{\MAX}-1)^2 \epsilon^2, \\
		\xi^{\DL}(\epsilon) &= \frac{1}{ \left( 1/\rho_{\MIN}^{\DL} + \gamma_{\MAX} + \gamma_{\MAX} (K_{\MAX}-1) \epsilon \right)^2 },
		\end{align}
	\end{subequations}
respectively, with $K_{\MAX} = \max\limits_{g,r} \abs{\cK_{\left(g,r\right)}}$, $\rho_{\MAX}^{\DL} = \max\limits_{g,r}\max\limits_{k \in \cK_{\left(g,r\right)}} \rho_k^{\DL}$, $\rho_{\MIN}^{\DL} = \min\limits_{g,r}\min\limits_{k \in \cK_{\left(g,r\right)}} \rho_k^{\DL}$, $\gamma_{\MAX} = \max\limits_{g,r}\max\limits_{k \in \cK_{\left(g,r\right)}} \gamma_k$, and $\gamma_{\MIN} = \min\limits_{g,r}\min\limits_{k \in \cK_{\left(g,r\right)}} \gamma_k$,
provided that the inner product of the channel direction vectors of the UTs scheduled over the same time-frequency resource blocks satisfies $\left\lvert \bv_k^H \bv_j \right\rvert \le \epsilon $ for $\forall k \ne j$ and $k,j\in\cK_{\left(g,r\right)}$.
Moreover, when $\epsilon \to 0$, the lowed bound of the DL achievable ergodic rate in \eqref{eq:prop_lb} asymptotically tends to be equal to the upper bound of the DL achievable ergodic rate in \eqref{eq:upp_rk}, i.e.,
\begin{equation}
\lim_{\epsilon \to 0} \upnot{R}{lb}_{\DL} = \oneon{G_{\mathrm{x}}G_{\mathrm{y}}}\sum_{g=0}^{G_{\mathrm{x}}-1} \sum_{r=0}^{G_{\mathrm{y}}-1} \sum_{k\in\cK_{\left(g,r\right)}} \expect{ \log_2 \left\{ 1 + \rho_k^{\DL}\left\lvert g_{k}^{\DL} \right\rvert^2  \right\} } =  \upnot{R}{ub}_{\DL}.
\end{equation}
\end{prop}
\begin{IEEEproof}
	Please refer to \appref{app:prop_dl_lowerbound}.
\end{IEEEproof}

\propref{prop:dl_lowerbound} shows that the proposed approach with SAUG and sCSI based ASLNR maximization DL precoder performed in each UT group is asymptotically optimal when $\left\lvert \bv_k^H \bv_j \right\rvert \to 0$. Note that this condition coincides with the upper bound achieving condition presented in \propref{prop:dl_op}. Therefore, when the number of satellite antennas $M$ and/or the number of scheduled UT groups is sufficiently large, the proposed approach is asymptotically optimal, which indicates the potential of adopting massive MIMO to serve a large number of UTs in LEO satellite communications. In addition, when the previously derived conditions are not rigorously satisfied (which is the usual case in practice), the proposed sCSI based precoder and receiver can mitigate the inter-user interference and further enhance the transmission performance for satellite communications.

\section{Simulation Results}\label{sec:sim_n}

In this section, we provide simulation results to evaluate the performance of the proposed massive MIMO transmission approach for LEO satellite communications. The major simulation setup parameters are listed as follows. The numbers of antennas equipped at the satellite side are set to be $\dnnot{M}{x}=\dnnot{M}{y}=16$ with half-wavelength antenna spacing in both the x- and y-axes.
The channel Rician factor is set to be $\kappa_{k}=\kappa=10\ \mathrm{dB}$, and the channel power is normalized as $\gamma_{k}=\dnnot{M}{x}\dnnot{M}{y}$ for all UT $k$. In addition, the channel space angles $\vartheta_{k,p}^{\mathrm{x}}$ and $\vartheta_{k,p}^{\mathrm{y}}$ are independently and uniformly distributed in the interval $[-1,1)$ for all UTs. The numbers of UT groups in the proposed SAUG approach are set to be equal for both x- and y-axes, i.e., $G_{\mathrm{x}}=G_{\mathrm{y}}=G$. The number of UTs to be grouped is set as $G^{2}M$.

Note that massive MIMO has not been applied to LEO satellite communications, and we consider and compare the following DL precoding and UL receiving approaches in the simulations:
\begin{itemize}
\item \textbf{IntF:} An ideal interference-free (IntF) case where the interference from other scheduled UTs over the same time and frequency resource is ``genie-aided'' eliminated will be considered as the performance upper bound.
\item \textbf{iCSI:} Relying on the iCSI, the SLNR maximization DL precoder in \eqref{eq:dl_slnr_k_pre} and the SINR maximization UL receiver in \eqref{eq:ul_sinr_k_rec} are adopted, with the assumption that the iCSI can be ``genie-aided'' obtained.
\item \textbf{sCSI:} The proposed sCSI based ASLNR maximization DL precoder and ASINR maximization UL receiver in \eqref{eq:dl_aslnr_k_pre_lb} and \eqref{eq:ul_asinr_k_rec_lb} are adopted, respectively.
\item \textbf{Fixed:} DFT based fixed DL precoding and UL receiving vectors in \eqref{eq:dl_dft} are adopted.
\end{itemize}

\begin{figure*}[!t]
\centering
\subfloat[DL]{\centering\includegraphics[width=0.48\textwidth]{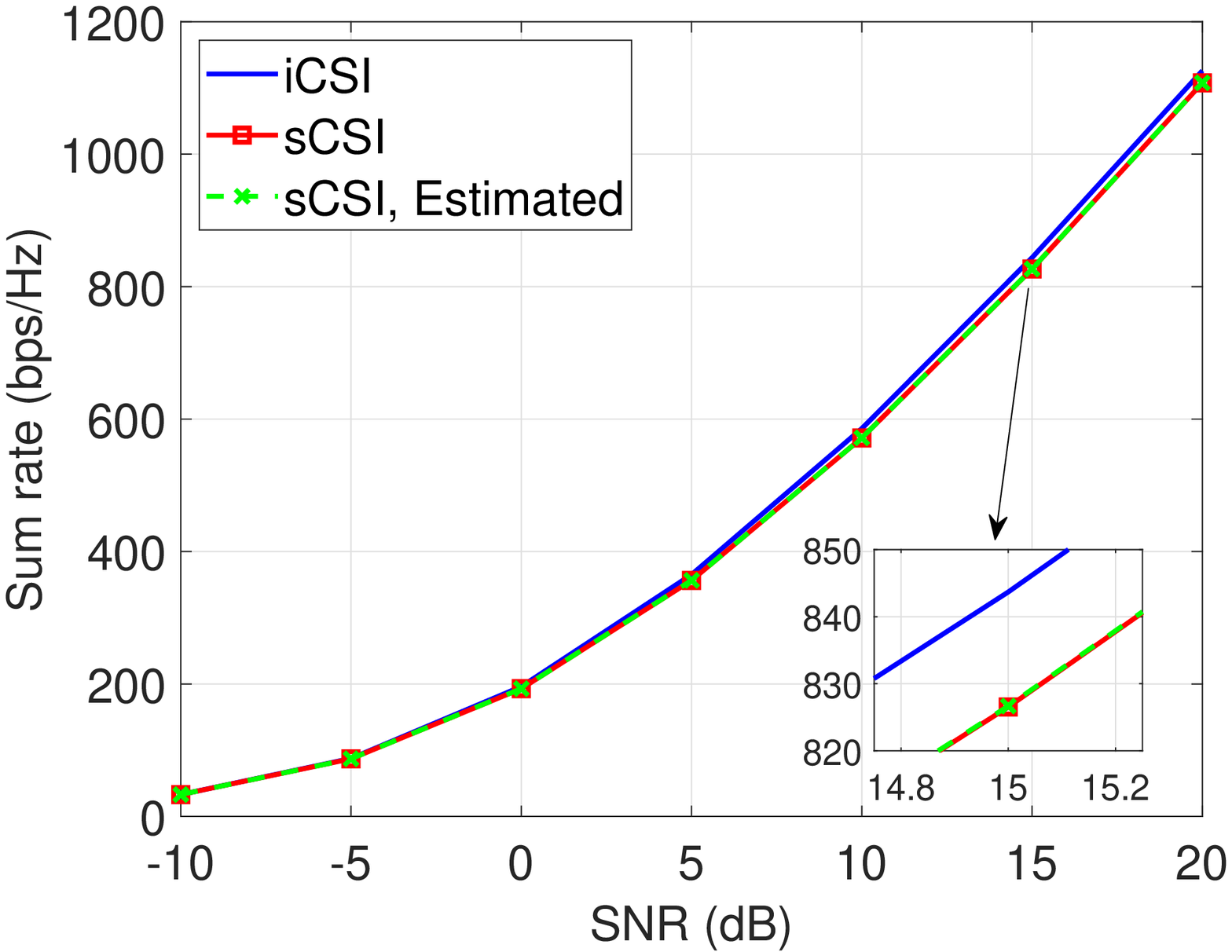}
\label{fig:dl_k10_isCSI_performance}}
\hfill
\subfloat[UL]{\centering\includegraphics[width=0.48\textwidth]{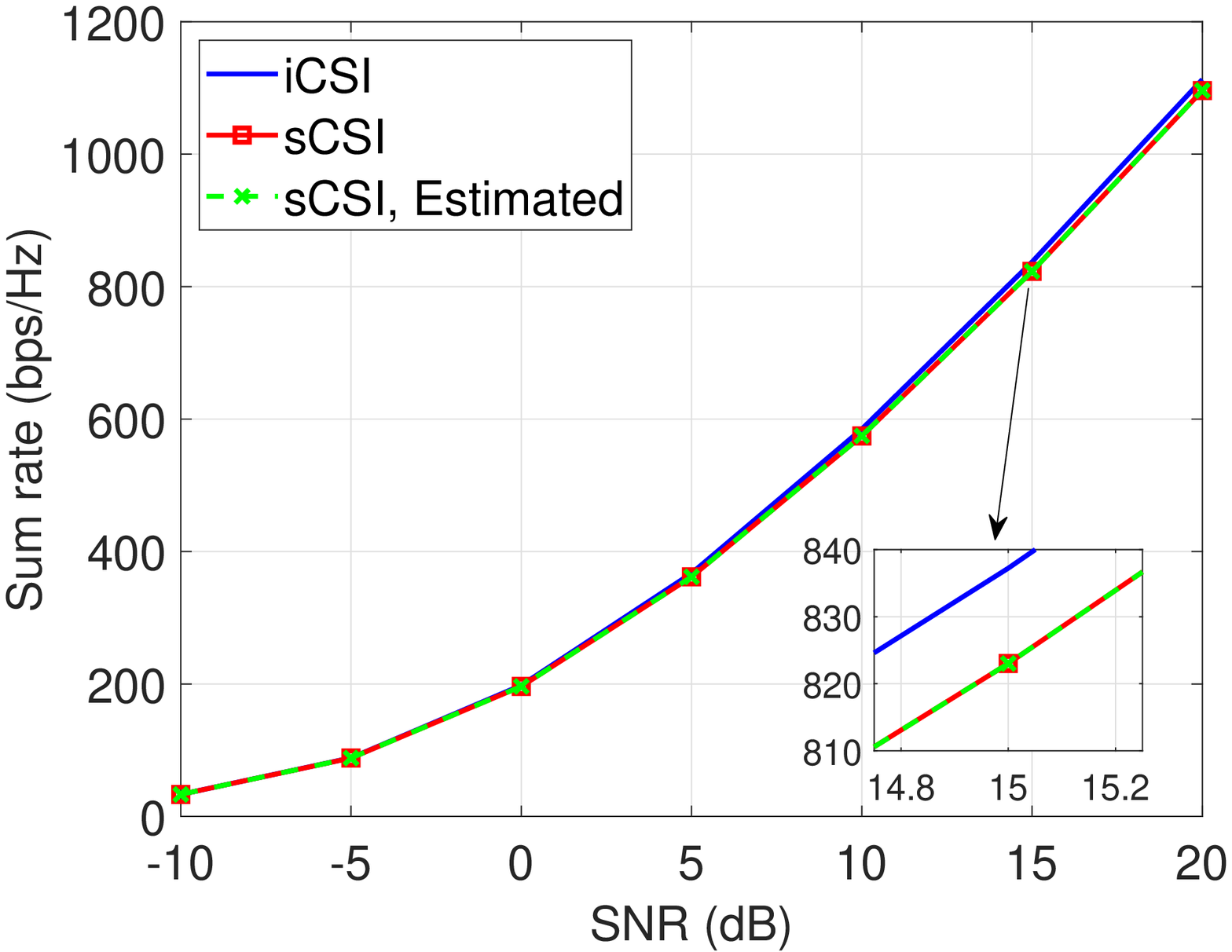}
\label{fig:ul_k10_isCSI_performance}}
\caption{Sum rate performance comparison between the proposed sCSI (using the true and the estimated sCSI that are obtained via averaging over 50 samples, respectively) and iCSI based precoding/receiving approaches.
}
\label{fig:gr_isCSI_Comparison}
\end{figure*}

In \figref{fig:gr_isCSI_Comparison}, we evaluate the performance of the proposed sCSI based precoding/receiving approaches, and compare them with the iCSI based ones where UTs are grouped using the proposed SAUG with $G=1$. We consider both cases that utilize the true and estimated sCSI that is obtained via averaging 50 samples. We can observe that in both UL and DL transmissions, the proposed sCSI based precoder and receivers exhibit almost identical performance as the iCSI based ones, while having significantly reduced computational overhead. In addition, the sum rate performance loss utilizing the estimated sCSI can be almost neglected.

\begin{figure*}[!t]
\centering
\subfloat[DL]{\centering\includegraphics[width=0.48\textwidth]{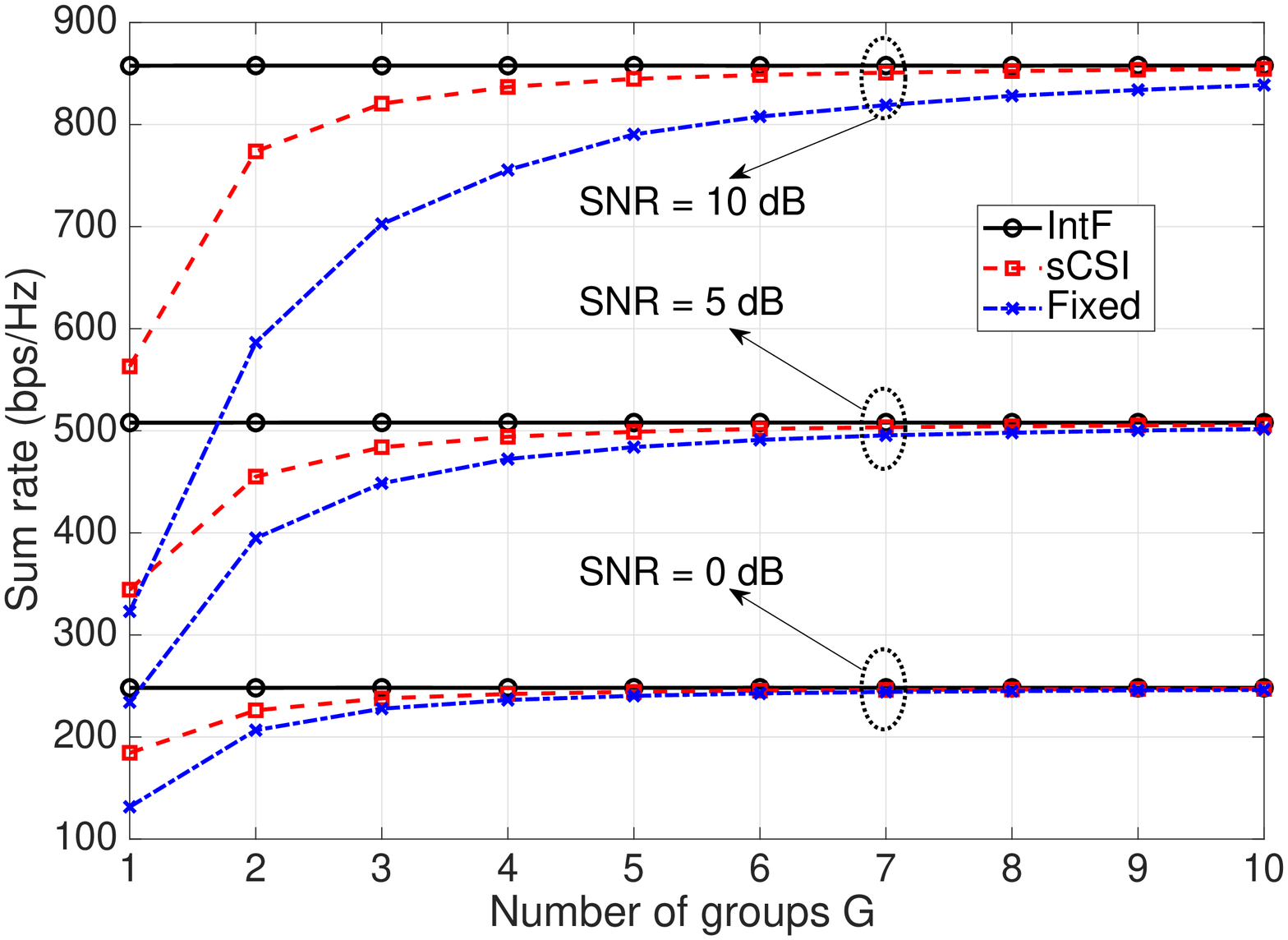}
\label{fig:dl_k10_gr_performance}}
\hfill
\subfloat[UL]{\centering\includegraphics[width=0.48\textwidth]{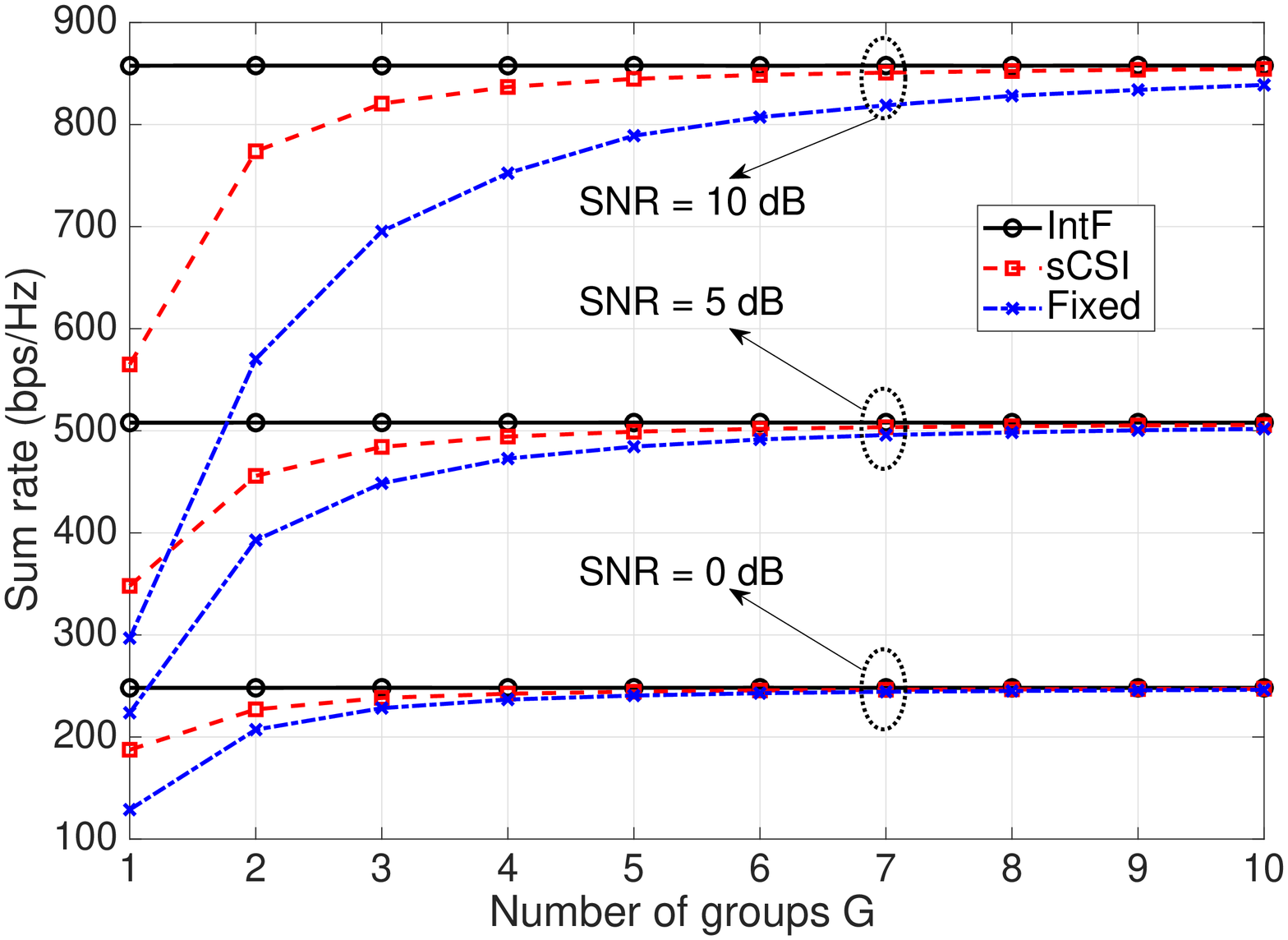}
\label{fig:ul_k10_gr_performance}}
\caption{Sum rate performance of SAUG with different transmission approaches versus the number of scheduled UT groups for different SNRs when FFR is adopted.}
\label{fig:gr_Comparison}
\end{figure*}

In \figref{fig:gr_Comparison}, we evaluate the performance of the proposed SAUG approach with different precoding/receiving approaches versus the number of scheduled groups $G$ when FFR is adopted across neighboring beams.
We can observe that the performance of the proposed sCSI based precoder/receiver can approach that of the interference-free scenario, especially in the case with a large number of scheduled groups, which demonstrates the asymptotic optimality of the proposed transmission approach.
In addition, the performance gap between the approach with fixed precoding/receiving vectors and the proposed sCSI based ones becomes smaller as the number of scheduled groups increases, especially in the low SNR regime, which indicates the near-optimality of the approach with fixed precoding/receiving vectors in the case where interference is not dominated.

\begin{figure*}[!t]
\subfloat[DL]{\centering\includegraphics[width=0.48\textwidth]{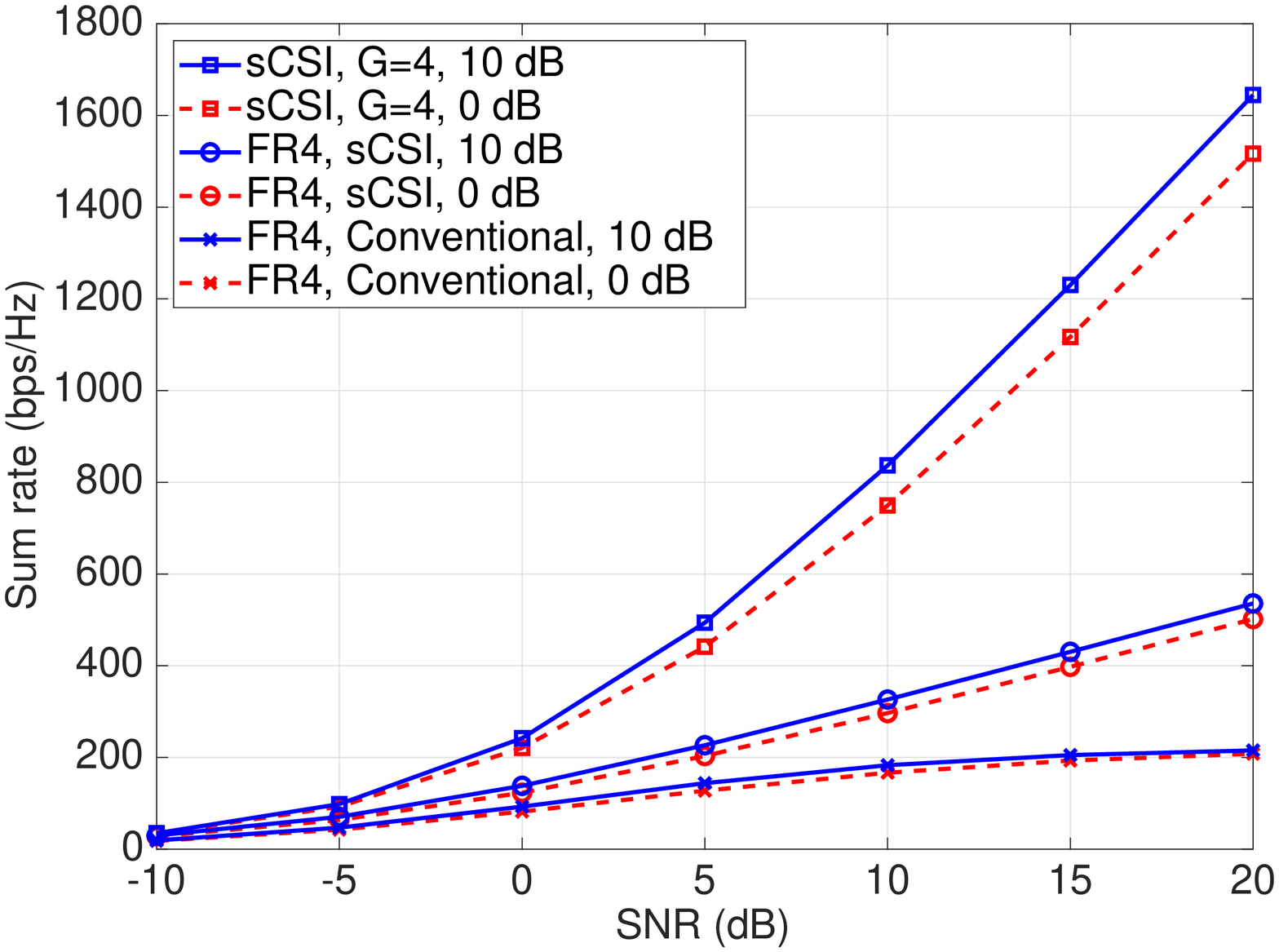}
\label{fig:ut_dl_k10_performance}}
\hfill
\subfloat[UL]{\centering\includegraphics[width=0.48\textwidth]{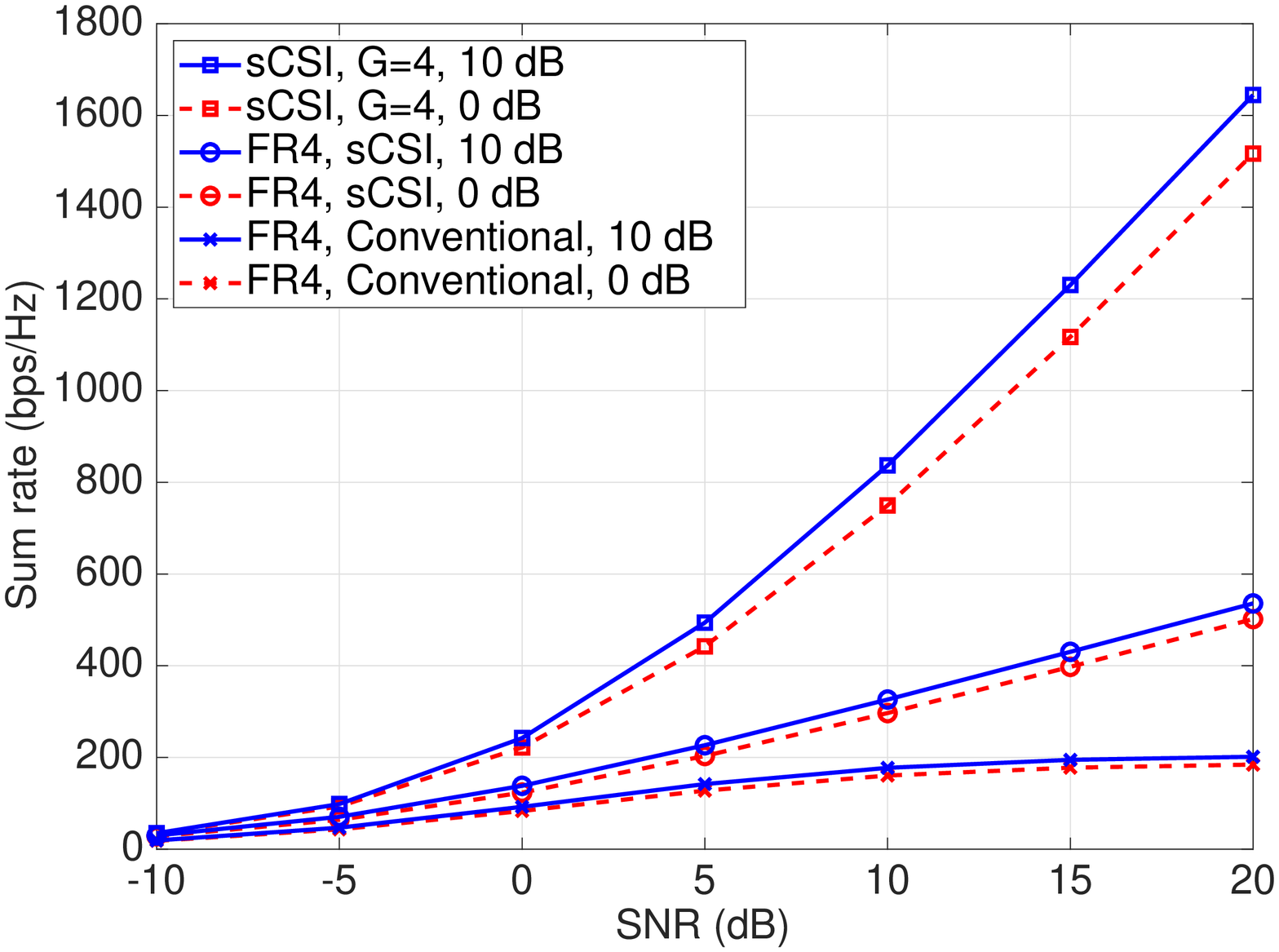}
\label{fig:ut_ul_k10_performance}}
\caption{Sum rate performance comparison between the proposed approach with FFR and the conventional FR4 approach under different Rician factors.}
\label{fig:ut_Comparison}
\end{figure*}

In \figref{fig:ut_Comparison}, the performance between the proposed transmission approach with FFR and the conventional FR4 approach is compared for different SNRs and channel Rician factors. Similarly as FFR, only one UT is scheduled per beam over the same time and frequency resource in FR4. Note that in the case of FR4, the UTs with the same color are a group of UTs performing transmission over the same time and frequency resource.
For FR4, we consider two transmission approaches where ``FR4, Conventional'' denotes the fixed precoder/receiver in \eqref{eq:dl_dft} and ``FR4, sCSI'' denotes the proposed sCSI based precoder/receiver in \eqref{eq:dl_aslnr_k_pre_lb}/\eqref{eq:ul_asinr_k_rec_lb} applied to the group of UTs over the same time and frequency resource for interference mitigation, respectively. We can observe that the proposed sCSI based precoder/receiver applied to FR4 show sum rate performance gains over the conventional FR4 approach. Moreover, with FFR across neighboring beams, the proposed sCSI based precoder/receiver combined with SAUG can provide significant sum rate performance gains over the conventional FR4 approach, especially in the cases with high SNRs and large Rician factors. Notably, for both UL and DL with an SNR of $20\;\mathrm{dB}$ and $\kappa=10\ \mathrm{dB}$, the proposed transmission approach with $G=4$ can provide about eight-folded sum rate performance gain over the conventional FR4 approach.

\section{Conclusion}\label{sec:conc}

In this paper, we have investigated massive MIMO transmission for LEO satellite communications exploiting sCSI with FFR. We first established the massive MIMO channel model for LEO satellite communications by taking into account the LEO satellite signal propagation properties and simplified the UL/DL
transmission designs via performing Doppler and delay compensations at UTs. Then, we developed the sCSI based DL precoder and UL receiver in closed-form, under the criteria of maximizing the ASLNR and the ASINR, respectively, and revealed the duality between them.
We further showed that the DL ASLNRs and UL ASINRs can reach their upper bounds provided that the channel direction vectors of the simultaneously served UTs are orthogonal, and proposed a space angle based user grouping (SAUG) approach motivated by this condition. Besides, we showed the asymptotic optimality of the proposed massive MIMO transmission approach exploiting sCSI. Simulation results showed that the proposed massive MIMO transmission scheme with FFR significantly enhances the data rate of LEO satellite communication systems. Notably, the proposed sCSI based precoder and receiver achieved the similar performance with the iCSI based ones that are often infeasible in practice. Future work includes detailed investigation on low complexity sCSI estimation, transmission designs for the cases with UTs using multiple antenna or directive antennas, low peak-to-average power ratio transmission signal design, and extension to the multiple LEO satellite communication systems, etc.

\appendices

\section{Proof of {\propref{prop:dl_op}}}\label{app:prop_dl_op}

We focus on the proof of the DL case and the proof of the UL case can be similarly obtained.
We first show the upper bound of $\upnot{\mathsf{ASLNR}}{max}_{k}$ in \eqref{eq:dl_aslnr_ub}.
From \eqref{eq:dl_aslnr_k_max}, we can obtain that $\upnot{\mathsf{ASLNR}}{max}_{k}$ with the proposed sCSI based precoder can be upper bounded by
\begin{align}\label{eq:dl_aslnr_k_max_con}
  \upnot{\mathsf{ASLNR}}{max}_{k}
  &=\frac{1}{1-\gamma_{k}\bv_{k}^{H}\left(\sum_{i}\gamma_{i}\bv_{i}\bv_{i}^{H}
  +\oneon{\upnot{\rho}{dl}_{k}}\bI_{M}\right)^{-1}\bv_{k}}-1 \ntb
  &\mathop{\leq}^{\mathrm{\left(a\right)}}
  \frac{1}{1-\gamma_{k}\bv_{k}^{H}\left(\gamma_{k}\bv_{k}\bv_{k}^{H}
  +\oneon{\upnot{\rho}{dl}_{k}}\bI_{M}\right)^{-1}\bv_{k}}-1 \ntb
  &\equab\upnot{\rho}{dl}_{k}\gamma_{k},
\end{align}
where (a) follows from that $\gamma_{i}\bv_{i}\bv_{i}^{H}$ is positive semidefinite for $\forall i$, and (b) follows from the Sherman-Morrison formula \cite[Eq. (15.2b)]{Seber08Matrix}.

We then show the achievability of the upper bound.
From the definition of $\bv_{k}$ in \eqref{eq:arr_resp_sing}, the condition given in \eqref{eq:dl_aslnr_k_max_con_ac} is equivalent to $\bv_{k}^{H}\bv_{i}=0$ for $\forall k\neq i$.
Thus, we can obtain that
\begin{align}\label{eq:dl_aslnr_ub_co}
  \left(\sum_{i}\gamma_{i}\bv_{i}\bv_{i}^{H}
  +\oneon{\upnot{\rho}{dl}_{k}}\bI_{M}\right)\bv_{k}\bv_{k}^{H}
  &=\left(\gamma_{k}\bv_{k}\bv_{k}^{H}
  +\oneon{\upnot{\rho}{dl}_{k}}\bI_{M}\right)\bv_{k}\bv_{k}^{H}\ntb
  &=\bv_{k}\bv_{k}^{H}\left(\gamma_{k}\bv_{k}\bv_{k}^{H}
  +\oneon{\upnot{\rho}{dl}_{k}}\bI_{M}\right),
\end{align}
which yields
\begin{align}\label{eq:dl_aslnr_ub_co2}
  \left(\sum_{i}\gamma_{i}\bv_{i}\bv_{i}^{H}
  +\oneon{\upnot{\rho}{dl}_{k}}\bI_{M}\right)^{-1}\bv_{k}\bv_{k}^{H}
  =\bv_{k}\bv_{k}^{H}\left(\gamma_{k}\bv_{k}\bv_{k}^{H}
  +\oneon{\upnot{\rho}{dl}_{k}}\bI_{M}\right)^{-1}.
\end{align}
Taking the traces of both sides of \eqref{eq:dl_aslnr_ub_co2}, we can further obtain
\begin{align}\label{eq:dl_aslnr_ub_co3}
  \bv_{k}^{H}\left(\sum_{i}\gamma_{i}\bv_{i}\bv_{i}^{H}
  +\oneon{\upnot{\rho}{dl}_{k}}\bI_{M}\right)^{-1}\bv_{k}
  =\bv_{k}^{H}\left(\gamma_{k}\bv_{k}\bv_{k}^{H}
  +\oneon{\upnot{\rho}{dl}_{k}}\bI_{M}\right)^{-1}\bv_{k}.
\end{align}
Thus, the inequality in (a) of \eqref{eq:dl_aslnr_k_max_con} can be obtained when the condition in \eqref{eq:dl_aslnr_k_max_con_ac} is satisfied. This concludes the proof.

\section{Proof of {\propref{prop:sp_rate_max_cond}}}\label{app:prop_dl_upbound}

The achievable ergodic rate in \eqref{eq:dlralsnr} can be upper bounded by
\begin{align}\label{eq:rate_max}
  \dnnot{R}{dl}
  &\mathop{\leq}^{(\mathrm{a})} \oneon{G_{\mathrm{x}}G_{\mathrm{y}}}\sum_{g=0}^{G_{\mathrm{x}}-1} \sum_{r=0}^{G_{\mathrm{y}}-1} \sum_{k\in\cK_{\left(g,r\right)}}\expect{\log_{2}\left\{1+
  \upnot{\rho}{dl}_{k}\sqrabs{\upnot{g}{dl}_{k}}\sqrabs{\bv_{k}^{T}\bb_k^{\ASLNR}}
  \right\}} \ntb
  &\mathop{\leq}^{(\mathrm{b})} \oneon{G_{\mathrm{x}}G_{\mathrm{y}}}\sum_{g=0}^{G_{\mathrm{x}}-1} \sum_{r=0}^{G_{\mathrm{y}}-1} \sum_{k\in\cK_{\left(g,r\right)}}\expect{\log_{2}\left\{1+
  \upnot{\rho}{dl}_{k}\sqrabs{\upnot{g}{dl}_{k}}
  \right\}},
\end{align}
where (a) follows from $\sqrabs{\cdot}\geq0$, and (b) follows from the Cauchy-Schwarz inequality.

We then examine the condition under which the upper bound in \eqref{eq:upp_rk} can be achieved. The inequality (a) in \eqref{eq:rate_max} becomes tight when $\bb_i^{\ASLNR}$ is orthogonal to $\left(\bv_{k}^\mathrm{x}\otimes\bv_{k}^\mathrm{y}\right)^{*}$ for $i\neq k$. In addition, as the equality in (b) can be achieved when $\bb_k^{\ASLNR}=\left(\bv_{k}^\mathrm{x}\otimes\bv_{k}^\mathrm{y}\right)^{*}$ \cite{Tulino06Capacity,Gao09Statistical}, we can obtain that for $\forall k\neq i\in\cK_{\left(g,r\right)}$, the channel direction vectors should satisfy $\left(\bv_{k}^\mathrm{x}\otimes\bv_{k}^\mathrm{y}\right)^{H}\left(\bv_{i}^\mathrm{x}\otimes\bv_{i}^\mathrm{y}\right)
=\left(\bv_{k}^\mathrm{x}\right)^{H}\bv_{i}^\mathrm{x}\left(\bv_{k}^\mathrm{y}\right)^{H}\bv_{i}^\mathrm{y}
=0$, i.e., $\left(\bv_{k}^\mathrm{x}\right)^{H}\bv_{i}^\mathrm{x}=0$ or $\left(\bv_{k}^\mathrm{y}\right)^{H}\bv_{i}^\mathrm{y}=0$.
This concludes the proof.

\section{Proof of {\propref{prop:dl_lowerbound}}}\label{app:prop_dl_lowerbound}

We first define some auxiliary variables for clarity of further proof. For notational brevity, we focus on a specific UT group, namely, the $\left(g,r\right)$th UT group $\cK_{\left(g,r\right)}$, and omit the group index as the UTs in different groups are scheduled over different time-frequency transmission resources.
For a given UT $k\in\cK_{\left(g,r\right)}$, we define $\udbdb_k^{\ASLNR} \triangleq \left(\sum_{i} \gamma_{i}\bv_{i}\bv_{i}^{H}
+\oneon{\upnot{\rho}{dl}_{k}}\bI_{M}\right)^{-1}\bv_{k}$. From \eqref{eq:dl_aslnr_k_pre_lb}, we can have
$\bb_k^{\ASLNR} = \left(  \udbdb_k^{\ASLNR} / \left\lVert \udbdb_k^{\ASLNR} \right\rVert  \right)^*$.
Then, the DL sum rate in \eqref{eq:dlralsnr} can be rewritten as
\begin{align}\label{eq:dlsrre}
R_{\DL}
&= \oneon{G_{\mathrm{x}}G_{\mathrm{y}}}\sum_{g=0}^{G_{\mathrm{x}}-1} \sum_{r=0}^{G_{\mathrm{y}}-1} \sum_{k\in\cK_{\left(g,r\right)}} \expect{ \log_2 \left\{ 1 + \frac{ \left\lvert g_{k}^{\DL} \right\rvert^2 \frac{ \left\lvert \bv_k^H \udbdb_k^{\ASLNR} \right\rvert^2 }{ \left\lVert \udbdb_k^{\ASLNR} \right\rVert^2  } q_k^{\DL} }{ \sum_{i\in\cK_{\left(g,r\right)}}^{i \ne k} \left\lvert g_{k}^{\DL} \right\rvert^2 \frac{ \left\lvert \bv_k^H \udbdb_i^{\ASLNR} \right\rvert^2 }{ \left\lVert \udbdb_i^{\ASLNR} \right\rVert^2  } q_i^{\DL} + \sigma_k^{\DL}  } \right\} }.
\end{align}
In order to obtain a lower bound of $R_{\DL}$ in \eqref{eq:dlsrre}, we provide an upper bound of $\left\lvert \bv_k^H \udbdb_i^{\ASLNR} \right\rvert^2$ for $\forall k\neq i$, a lower bound of $\left\lVert \udbdb_k^{\ASLNR} \right\rVert^2$, and a lower bound of $\left\lvert \bv_k^H \udbdb_k^{\ASLNR} \right\rvert^2 / \left\lVert \udbdb_k^{\ASLNR} \right\rVert^2 $, respectively, in the following.

\subsection{Upper Bound of $\left\lvert \bv_k^H \udbdb_i^{\ASLNR} \right\rvert^2$ for $\forall k\neq i$}

Denote by $K\triangleq\abs{\cK_{\left(g,r\right)}}$, $\bV \triangleq \left[ \bv_1,\dots,\bv_{K} \right] $, and $\bdGamma \triangleq \diag{ \left[\gamma_1,\dots,\gamma_{K} \right]^{T}}$. Then,
$ \bv_k^H \udbdb_i^{\ASLNR} $
can be expressed by the $(k,i)$th element of the following matrix
\begin{align}
\bA
&\triangleq \bV^H \left( \bV \bdGamma \bV^H + \oneon{\upnot{\rho}{dl}_{k}}\bI_{M} \right)^{-1} \bV
\equaa \bdGamma^{-1} - \oneon{\upnot{\rho}{dl}_{k}} \bdGamma^{-1} \left( \oneon{\upnot{\rho}{dl}_{k}} \bdGamma^{-1} + \bV^H \bV \right)^{-1} \bdGamma^{-1},
\end{align}
where (a) follows from the matrix inversion lemma. Let $\bB \triangleq \oneon{\upnot{\rho}{dl}_{k}} \bdGamma^{-1} + \bV^H \bV$. Then, $ \bv_k^H \udbdb_i^{\ASLNR} $ can be further written as
\begin{align}\label{eq:vkbia}
\bv_k^H \udbdb_i^{\ASLNR} =\vecele{\bA}{k,i}=
\begin{cases}
-\frac{1}{\upnot{\rho}{dl}_{k} \gamma_i \gamma_k } \left[ \bB^{-1} \right]_{k,i} ,& \text{ if } i \ne k \\
\frac{1}{\gamma_k} - \frac{1}{\upnot{\rho}{dl}_{k} \gamma_k^2 } \left[ \bB^{-1} \right]_{k,k} ,& \text{ if } i = k
\end{cases}.
\end{align}

Denote by $b_{k,j}$ the $\left(k,j\right)$th element of $\bB$, and $D_k'\left( \bB \right) \triangleq \sum_{j \ne k } \left\lvert b_{k,j} \right\rvert = \sum_{j \ne k } \left\lvert \bv_k^H \bv_j \right\rvert$. Then, according to Ger\v{s}gorin disc theorem \cite[Theorem 6.1.1]{Horn12Matrix}, for an arbitrary eigenvalue $\lambda$ of $\bB$, there exists an integer $ 1 \le p \le K$ such that
\begin{align}\label{Gersgorin_disc_inequality}
\left\lvert \lambda - b_{p,p} \right\rvert \le D_p'\left( \bB \right) \mathop{\leq}^{\mathrm{\left(a\right)}} \left(K-1\right)\epsilon
\mathop{\leq}^{\mathrm{\left(b\right)}}\left(\dnnot{K}{max}-1\right)\epsilon,
\end{align}
where (a) follows from $\left\lvert \bv_p^H \bv_j \right\rvert \le \epsilon$ for all $j \ne p$, and (b) follows from $K\leq \dnnot{K}{max}$.
Denote by $\lambda_{\MAX}$ and $\lambda_{\MIN}$ the largest and smallest eigenvalues of $\mathbf{B}$, respectively. For $\forall i \ne k$, we can have the following inequality
\begin{align}\label{B_UL_inej_upper_bound}
\left\lvert \left[ \bB^{-1} \right]_{k,i} \right\rvert
&= \left\lvert \mathbf{e}_k^T \bB^{-1} \mathbf{e}_i \right\rvert
\mathop{\leq}^{\mathrm{\left(a\right)}}  \frac{ 1/\lambda_{\MIN} - 1/\lambda_{\MAX} }{ 1/\lambda_{\MIN} + 1/\lambda_{\MAX} }  \sqrt{ \mathbf{e}_k^T \bB^{-1} \mathbf{e}_k } \sqrt{ \mathbf{e}_i^T \bB^{-1} \mathbf{e}_i } \ntb
&= \frac{ \lambda_{\MAX} - \lambda_{\MIN} }{ \lambda_{\MAX} + \lambda_{\MIN } } \sqrt{ \left[ \bB^{-1} \right]_{k,k} } \sqrt{ \left[ \bB^{-1} \right]_{i,i} }
\mathop{\leq}^{\mathrm{\left(b\right)}} \frac{ 2\left(\dnnot{K}{max}-1\right) \epsilon }{ \lambda_{\MAX} + \lambda_{\MIN} } \sqrt{ \left[ \bB^{-1} \right]_{k,k} } \sqrt{ \left[ \bB^{-1} \right]_{i,i} } \ntb
&\mathop{\leq}^{\mathrm{\left(c\right)}}  \rho_{\MAX}^{\DL} \gamma_{\MAX}\left(\dnnot{K}{max}-1\right) \epsilon  \sqrt{ \left[ \bB^{-1} \right]_{k,k} } \sqrt{ \left[ \bB^{-1} \right]_{i,i} } \ntb
&\mathop{\leq}^{\mathrm{\left(d\right)}} \left( \rho_{\MAX}^{\DL}\gamma_{\MAX} \right)^2 \left(\dnnot{K}{max}-1\right) \epsilon,
\end{align}
where $\mathbf{e}_k$ is the $k$th column of identity matrix, (a) follows from Wielandt's inequality \cite[Eq. (7.4.12.2)]{Horn12Matrix}, (b) follows from \eqref{Gersgorin_disc_inequality}, (c) follows from Weyl's inequality \cite[Corollary 4.3.15]{Horn12Matrix}, and (d) follows from Rayleigh quotient theorem \cite[Theorem 4.2.2(c)]{Horn12Matrix}. From \eqref{eq:vkbia} and the inequality in \eqref{B_UL_inej_upper_bound}, we can obtain
\begin{align}\label{eq:bvkbbi}
\left\lvert \bv_k^H \udbdb_i^{\ASLNR} \right\rvert^2
\le  \frac{\left( \rho_{\MAX}^{\DL}\gamma_{\MAX} \right)^4}{ \left( \rho_{k}^{\DL} \gamma_i \gamma_k \right)^2 }  \left(\dnnot{K}{max}-1\right)^2 \epsilon^2 \triangleq \beta_{k,i}^{\DL} (\epsilon), \quad \forall i \ne k.
\end{align}

\subsection{Lower Bound of $\left\lVert \udbdb_k^{\ASLNR} \right\rVert^2$}

Defining $\zeta_{\MAX}(\bX)$ as the maximum eigenvalue of matrix $\bX$. Then, a lower bound of $\left\lVert \udbdb_k^{\ASLNR} \right\rVert^2$ can be obtained as
\begin{align}\label{udbdb_norm_lower_bound}
\left\lVert \udbdb_k^{\ASLNR} \right\rVert^2
&= \bv_k^H \left( \bV \bdGamma \bV^H + \oneon{\upnot{\rho}{dl}_{k}}\bI_{M} \right)^{-2} \bv_k
\mathop{\geq}^{\mathrm{\left(a\right)}} \frac{1}{ \zeta_{\MAX}^2 \left( \bV \bdGamma \bV^H + \oneon{\upnot{\rho}{dl}_{k}}\bI_{M} \right) } \ntb
&\mathop{\geq}^{\mathrm{\left(b\right)}} \frac{1}{ \left(  1/\rho_k^{\DL} +  \zeta_{\MAX}\left( \bV \bdGamma \bV^H \right) \right)^2 }
\mathop{\geq}^{\mathrm{\left(c\right)}} \frac{1}{ \left( 1/\rho_{\MIN}^{\DL} + \gamma_{\MAX} + \gamma_{\MAX} (\dnnot{K}{max}-1) \epsilon \right)^2 } \triangleq \xi^{\DL} \left( \epsilon \right),
\end{align}
where (a) follows from Rayleigh quotient theorem \cite[Theorem 4.2.2(c)]{Horn12Matrix}, (b) follows from Weyl's inequality \cite[Corollary 4.3.15]{Horn12Matrix}, and (c) holds by applying Ger\v{s}gorin disc theorem \cite[Theorem 6.1.1]{Horn12Matrix} to matrix $\bV \bdGamma \bV^H$.

\subsection{Lower Bound of $\left\lvert \bv_k^H \udbdb_k^{\ASLNR} \right\rvert^2 / \left\lVert \udbdb_k^{\ASLNR} \right\rVert^2 $}

Before proceeding, we first present some preliminary results. An upper bound of $\left\lVert \udbdb_k^{\ASLNR} \right\rVert^2$ can be obtained as
\begin{align}\label{uk_norm_upper_bound}
\left\lVert \udbdb_k^{\ASLNR} \right\rVert^2
&=\left\lVert \left( \bV \bdGamma \bV^H + \oneon{\upnot{\rho}{dl}_{k}}\bI_{M} \right)^{-1} \bV \be_k \right\rVert^2
\equaa \left\lVert \bV \bB^{-1} \bdGamma^{-1} \be_k \right\rVert^2
= \be_k^T \bdGamma^{-1}  \bB^{-1} \bV^H \bV \bB^{-1} \bdGamma^{-1} \be_k \ntb
&\equab \frac{1}{\gamma_k^2} \left[ \bB^{-1} \left( \bB - \frac{1}{\rho_k^{\DL}} \bdGamma^{-1} \right) \bB^{-1}  \right]_{k,k}
= \frac{1}{\gamma_k^2} \left( \left[ \bB^{-1} \right]_{k,k} - \frac{1}{\rho_k^{\DL}} \left[ \bB^{-1} \bdGamma^{-1} \bB^{-1}  \right]_{k,k}  \right)  \ntb
&= \frac{1}{\gamma_k^2} \left( \left[ \bB^{-1} \right]_{k,k} - \sum_{j=1}^{K} \frac{1}{\rho_k^{\DL} \gamma_j } \left\lvert \left[ \bB^{-1} \right]_{k,j} \right\rvert^2  \right) \ntb
&\leq \frac{1}{\gamma_k^2}  \left[ \bB^{-1} \right]_{k,k} \left( 1 - \frac{1}{\rho_k^{\DL} \gamma_k } \left[ \bB^{-1} \right]_{k,k} \right) \triangleq \eta_k^{\DL},
\end{align}
where (a) follows from the matrix inversion lemma, and (b) follows from the definition of $\bdGamma$ and $\bB$.
In addition, $\left[ \bB^{-1} \right]_{k,k}$ can be lower bounded by
\begin{align}\label{eq:kaidle}
\left[ \bB^{-1} \right]_{k,k}
\mathop{\geq}^{\mathrm{\left(a\right)}} \frac{1}{\lambda_{\MAX}}
\mathop{\geq}^{\mathrm{\left(b\right)}} \frac{1}{b_{p,p}+(\dnnot{K}{max}-1)\epsilon}
\ge \frac{1}{ \frac{1}{\rho_{\MIN}^{\DL} \gamma_{\MIN} } + 1 + (\dnnot{K}{max}-1) \epsilon } \triangleq \chi^{\DL}\left( \epsilon \right),
\end{align}
where (a) follows from Rayleigh quotient theorem \cite[Theorem 4.2.2(c)]{Horn12Matrix}, (b) follows from Ger\v{s}gorin disc theorem \cite[Theorem 6.1.1]{Horn12Matrix}.
Moreover, the following inequality can be obtained
\begin{align}\label{one_akk_eta_required_inequality}
1
&= \left\lvert \left[ \bB \bB^{-1} \right]_{k,k}  \right\rvert
= \left\lvert \sum_{j=1}^{K} b_{k,j} \left[ \bB^{-1} \right]_{j,k}  \right\rvert \ntb
&\mathop{\geq}^{\mathrm{\left(a\right)}}   b_{k,k}  \left[ \bB^{-1} \right]_{k,k}  - \sum_{j \ne k} \left\lvert  b_{k,j} \right\rvert \left\lvert  \left[ \bB^{-1} \right]_{j,k} \right\rvert
\mathop{\geq}^{\mathrm{\left(b\right)}}   b_{k,k}  \left[ \bB^{-1} \right]_{k,k} -  \sum_{j \ne k} \left( \rho_{\MAX}^{\DL} \gamma_{\MAX} \right)^2 (\dnnot{K}{max}-1) \epsilon^2   \ntb
&=  \left(  \frac{1}{\rho_k^{\DL} \gamma_k } + 1 \right) \left[ \bB^{-1} \right]_{k,k} - \left( \rho_{\MAX}^{\DL}\gamma_{\MAX} \right)^2 (\dnnot{K}{max}-1)^2 \epsilon^2,
\end{align}
where (a) follows from the triangle inequality, and (b) follows from $\left\lvert b_{k,j} \right\rvert=\abs{\bv_{k}^{H}\bv_{j}} \le \epsilon$ for $\forall j \ne k$ and the inequality in \eqref{B_UL_inej_upper_bound}. Then, we can obtain
\begin{align}\label{eq:bin2}
\frac{ 1 }{ \left[ \bB^{-1} \right]_{k,k} } - \frac{1}{\upnot{\rho}{dl}_{k} \gamma_k }
&\mathop{\geq}^{\mathrm{\left(a\right)}}
1-\oneon{ \left[ \bB^{-1} \right]_{k,k} }\left( \rho_{\MAX}^{\DL}\gamma_{\MAX} \right)^2 (\dnnot{K}{max}-1)^2 \epsilon^2 \ntb
&\mathop{\geq}^{\mathrm{\left(b\right)}}
1 - \frac{ \left( \rho_{\MAX}^{\DL}\gamma_{\MAX} \right)^2 (\dnnot{K}{max}-1)^2 \epsilon^2 }{ \chi^{\DL}(\epsilon)  }
\triangleq 1 - \delta^{\DL} \left( \epsilon \right),
\end{align}
where (a) follows from \eqref{one_akk_eta_required_inequality}, and (b) follows from \eqref{eq:kaidle}.

Consequently, $\left\lvert \bv_k^H \udbdb_k^{\ASLNR} \right\rvert^2 / \left\lVert \udbdb_k^{\ASLNR} \right\rVert^2 $ can be lower bounded by
\begin{align}\label{akk_eta}
\frac{\left\lvert \bv_k^H \udbdb_k^{\ASLNR} \right\rvert^2  }{\left\lVert \udbdb_k^{\ASLNR} \right\rVert^2}
&\mathop{\geq}^{\mathrm{\left(a\right)}}
\frac{ \left\lvert \bv_k^H \udbdb_k^{\ASLNR} \right\rvert^{2} }{ \eta_k^{\DL} }
\equab \frac{ \left( 1 - \frac{1}{\upnot{\rho}{dl}_{k} \gamma_k } \left[ \bB^{-1} \right]_{k,k} \right)^2 }{ \gamma_k^2 \eta_k^{\DL} }
\equac  \frac{ 1 }{ \left[ \bB^{-1} \right]_{k,k} } - \frac{1}{\upnot{\rho}{dl}_{k} \gamma_k }
\mathop{\geq}^{\mathrm{\left(d\right)}}
1 - \delta^{\DL} \left( \epsilon \right),
\end{align}
where (a) follows from the inequality in \eqref{uk_norm_upper_bound}, (b) follows from \eqref{eq:vkbia}, (c) follows from the definition of $\eta_k^{\DL}$ in \eqref{uk_norm_upper_bound}, and (d) follows from \eqref{eq:bin2}.

Combining \eqref{eq:dlsrre}, \eqref{eq:bvkbbi}, \eqref{udbdb_norm_lower_bound}, and \eqref{akk_eta}, we can obtain a lower bound of $R_{\DL}$ as
\begin{align}\label{sumrate_DL_lower_bound}
R_{\DL}
&\ge \oneon{G_{\mathrm{x}}G_{\mathrm{y}}}\sum_{g=0}^{G_{\mathrm{x}}-1} \sum_{r=0}^{G_{\mathrm{y}}-1} \sum_{k\in\cK_{\left(g,r\right)}} \expect{ \log_2 \left\{ 1 + \frac{ \left\lvert g_{k}^{\DL} \right\rvert^2 \left( 1 - \delta^{\DL} \left( \epsilon \right) \right) q_k^{\DL} }{ \sum_{i\in\cK_{\left(g,r\right)}}^{i \ne k} \left\lvert g_{k}^{\DL} \right\rvert^2 \frac{  \beta_{k,i}^{\DL}(\epsilon) }{ \xi^{\DL}(\epsilon)  } q_i^{\DL} + \sigma_k^{\DL}  } \right\} } \triangleq \upnot{R}{lb}_{\DL}.
\end{align}
Note that $\lim_{\epsilon \rightarrow 0} \delta^{\DL} \left( \epsilon \right) = 0$, $\lim_{\epsilon \rightarrow 0} \beta_{k,i}^{\DL}(\epsilon) = 0$, and $\xi^{\DL}(\epsilon)>0$, thus we can obtain $\lim_{\epsilon \to 0} \upnot{R}{lb}_{\DL} = \upnot{R}{ub}_{\DL}$. This concludes the proof.

%\bibliographystyle{IEEEtran}
%\bibliography{Refabrv_20181123,References_20190524}


\end{document}